\documentclass{article}
\usepackage{authblk}
\usepackage[symbol]{footmisc}
\usepackage{amssymb}
\usepackage{amsmath} 
\usepackage{bbold}
\usepackage{enumitem}
\usepackage{booktabs}
\usepackage{threeparttable}
\usepackage{tikz}
\usetikzlibrary{positioning, arrows.meta}
\usepackage[section]{placeins}
\usepackage[english]{babel}
\usepackage[letterpaper,top=2cm,bottom=2cm,left=3cm,right=3cm,marginparwidth=1.75cm]{geometry}
\usepackage{amsmath}
\usepackage{graphicx}
\usepackage[colorlinks=true, allcolors=blue]{hyperref}

\newcommand{\V}{\mathbb{V}}
\newcommand{\E}{\mathbb{E}}

\newcommand{\EIF}{\mathbb{EIF}}
\newcommand{\ATT}{\textrm{ATT}}
\newcommand{\ATE}{\textrm{ATE}}
\newcommand{\Eq}[1]{Eq.~\eqref{#1}}
\newcommand{\Eqs}[2]{Eqs.~\eqref{#1} and~\eqref{#2}}
\newcommand{\Sec}[1]{Sec.~\ref{#1}}
\newcommand{\Secs}[2]{Secs.~\ref{#1} and~\ref{#2}}
\newcommand{\Fig}[1]{Fig.~\ref{#1}}

\title{Digital Twins as Synthetic Controls in Single-Arm Trials}
\author{Daniele Bertolini}
\author{Franklin Fuller}
\author{Aaron M. Smith}
\author{Jonathan R. Walsh}
\author{Run Zhuang}
\affil{Unlearn.AI, Inc., San Francisco, CA}

\begin{document}
\maketitle

\begin{abstract}
Single-arm trials are an important study design for evaluating drug efficacy and safety without enrolling patients into a control arm. Although they do not provide the gold-standard evidence of randomized controlled trials, they are increasingly used in clinical development as they offer an efficient, ethical, and practical alternative. A wide variety of approaches can be used to construct control comparators and estimate treatment effects, from fixed comparators informed by clinical knowledge to data-based and model-based patient-level comparators, also known as synthetic controls. Powerful and flexible machine learning models can allow outcome-model-based synthetic controls to overcome key limitations of direct data-based approaches, yield more robust estimates of treatment effects, and provide a principled way to incorporate corrections or encode additional assumptions when external data are not directly comparable. In this work, we argue that outcome-model-based synthetic control arms are an important tool for single-arm trials. We focus on digital twins, personalized predictions of disease progression generated from machine learning models trained on historical datasets, which naturally leverage these flexible approaches. We review doubly robust estimators, present power and sample size formulas, and discuss trade-offs in selecting historical data for training and analysis. We also outline practical considerations for deploying digital twins within the framework of recent FDA draft guidance on the use of artificial intelligence in drug development. Finally, we reanalyze data from trials in amyotrophic lateral sclerosis and Huntington’s disease to demonstrate the proposed methods.
\end{abstract}

\section{Introduction}
\label{sec:introduction}
Modern drug development faces substantial challenges, including rising clinical trial costs, competition for patients, slow recruitment, and increasing pressure to accelerate innovation. In this context, single-arm trials have emerged as an attractive study design, particularly in early clinical development, where rapid signal detection and efficient decision-making are critical. They are also used in settings such as accelerated approval pathways, open-label extensions, and post-approval studies, and may be considered when randomized controlled trials (RCTs) are infeasible or impractical. Compared to RCTs, single-arm trials do not require enrolling a concurrent control group, which can substantially reduce the total number of participants needed and shorten study timelines. This makes them a valuable tool for early go/no-go decisions and prioritization within a development program.

A central challenge in single-arm trials is the estimation of treatment effects in the absence of a concurrent control arm. As in any observational setting, identification of the estimand must be demonstrated given the available data. Conditional on identifiability—typically requiring that all relevant confounders are measured—a key problem is the appropriate adjustment for these confounders. This is particularly critical in single-arm trials, where inadequate adjustment can directly induce bias in treatment effect estimates, in contrast to RCTs where randomization ensures unbiasedness and covariate adjustment primarily improves efficiency. Addressing this challenge requires constructing a comparator using external data and accounting for differences between the trial population and the external sources. Existing approaches typically rely on direct comparisons, through matched or weighted external controls, most commonly via propensity score–based methods. While widely used, these approaches can struggle when external data are heterogeneous or incomplete, and often require restricting attention to subsets of the data where sufficient overlap holds, or else risk producing biased or unstable estimates.

Outcome-model-based approaches provide a complementary alternative. Rather than relying solely on direct comparisons to matched or weighted external controls, they model outcomes under a defined standard of care to estimate counterfactual responses. This enables the use of broader and more heterogeneous data sources and is particularly powerful when combined with flexible machine learning models that capture complex relationships between covariates and outcomes. Moreover, model-based approaches provide a natural framework for incorporating additional assumptions or corrections when identifiability is tenuous or when the available data are not directly comparable—for example, in the presence of covariate shift or structured missingness. Finally, outcome-model-based approaches can also be combined with propensity-based methods to improve robustness.

In this work, we focus on synthetic control arms constructed using outcome models, with particular emphasis on digital twins. Because terminology varies substantially across the community, we use “external controls” as an umbrella term for any comparator constructed from data outside the study, and “synthetic controls” to refer to patient-level controls constructed to reflect the characteristics of the specific population of interest, rather than constant outcomes such as population-level averages from past clinical trials. A digital twin is a patient-level model that comprehensively predicts future outcomes under a predefined standard of care from baseline characteristics. Such models can be trained on diverse data sources with varying levels of heterogeneity, granularity, modality, and missingness, and can leverage information across related datasets through modern machine learning techniques.  The goal of the digital twin is to capture as complete of a picture of patient outcomes as the data supports. Importantly, they provide individualized counterfactual predictions, aligning directly with the goal of constructing an external comparator by conditioning on each patient’s observed characteristics. Because they model patient trajectories over time, they naturally support comparisons across multiple endpoints and timepoints within a unified framework.

A potential limitation of using machine learning models to predict counterfactual outcomes is that naïvely using their predictions to estimate treatment effects can introduce bias. Even when outcome models are well-specified and asymptotically consistent, finite-sample bias—often induced by regularization—may not decay sufficiently quickly to be negligible. Doubly robust estimators address this issue by combining outcome models with propensity-based adjustments to debias predictions and relax the convergence rate requirements to practically attainable levels. These estimators yield consistent treatment effect estimates when either the outcome model or the propensity model is correctly specified, enable valid statistical inference, and achieve optimal asymptotic efficiency within the class of regular and asymptotically linear estimators. Among these, augmented inverse probability weighting (AIPW) provides a practical and well-understood estimator for common estimands. Despite these advantages of model-based approaches, most single-arm trials today either do not construct formal synthetic control arms or rely primarily on propensity score–based methods. As flexible outcome models become increasingly available, model-based approaches—particularly those based on digital twins—offer an opportunity to improve both robustness and efficiency in treatment effect estimation.

The contributions of this work are threefold. First, we review and discuss advantages of digital twin approaches in the context of single-arm trials. Second, after reviewing properties of common treatment effect estimators, we present new results on power and sample size calculations for AIPW estimators. Third, we discuss practical considerations for the selection and use of external data, discuss digital twins in single-arm trials in relation to the recent FDA draft guidance on artificial intelligence, and illustrate applications through case studies in amyotrophic lateral sclerosis (ALS) and Huntington’s disease (HD).

The remainder of the paper is organized as follows. We begin by reviewing the definition of average treatment effect on the treated and indentifiability conditions in \Sec{sec:assumptions} and strategies for choosing external comparators in \Sec{sec:strategy}. We then discuss common estimators and their properties in \Sec{sec:estimators} and summarize expected advantages of digital twin models for single-arm trials in \Sec{sec:digital-twins}. We discuss power and sample size computations for outcome-model-based estimators in \Sec{sec:power} and review data selection and the FDA guidance in \Secs{sec:external-data}{sec:fda-guidance}. We discuss case studies in \Sec{sec:case-studies} and conclusions in \Sec{sec:discussion}. Derivations and additional material are collected in the Appendices.

Readers focused on the
practical aspects of designing and interpreting single-arm trials,
such as how to choose an external comparator, when digital twins
offer practical advantages, and the worked examples, may concentrate
on \Sec{sec:strategy} and \Sec{sec:digital-twins}, and the case studies in \Sec{sec:case-studies}.
Optionally, they may also review \Secs{sec:external-data}{sec:fda-guidance} on data
selection and alignment with the FDA framework. Readers interested in
the statistical and methodological details, including formal
definitions and asymptotic properties of common estimators,
derivations of efficient influence functions and asymptotic
variances, and power and sample size calculations for
outcome-model--based estimators, may instead focus on the overview of
estimators in \Sec{sec:estimators} together with the corresponding
derivations in the Appendices, \Sec{sec:digital-twins}, the power
computations in \Sec{sec:power}, and the case studies in
\Sec{sec:case-studies}.

\section{Treatment Effect Definition and Identifiability}
\label{sec:assumptions}
In a randomized controlled trial (RCT), a standard measure of efficacy is the average treatment effect (ATE),
\[
\tau_{\ATE} = \E[Y(1) - Y(0)],
\]
where $Y(1)$ and $Y(0)$ denote the potential outcomes under treatment and control, respectively. Randomization ensures that treatment assignment is independent of patient characteristics, so observed differences in outcomes between treatment groups can be attributed to the treatment itself, with covariate adjustment primarily used to improve precision.
In a single-arm trial, only treated outcomes are observed, which shifts the focus to the average treatment effect on the treated (ATT),
\begin{equation}
\label{eq:att-def}
    \tau_{\ATT} = \E[Y(1) - Y(0) \mid A = 1],
\end{equation}
where $A \in \{0,1\}$ is a binary treatment indicator. In this setting, outcomes under control for treated patients, $Y(0) \mid A = 1$, are not observed and must be inferred from external data. Estimation of the ATT therefore relies on assumptions about the comparability of treated patients to external data sources, with confounders playing a central role. A confounder is a variable associated with both treatment assignment, here membership in the treated group versus the external control, and the outcome of interest. Examples of potential confounders in a trial could include baseline disease severity, demographic information, comorbidity burden, geographic location, medical history, or biomarker status, when these factors are prognostic of disease progression and differ between the trial and external control populations. If left unadjusted, a confounder distorts the apparent treatment effect because differences in outcomes between treated and untreated groups partly reflect the confounder rather than the treatment itself. In a randomized controlled trial, causal treatment effects can be estimated directly because randomization ensures that both measured and unmeasured confounders are balanced across treatment groups on average, thereby eliminating potential bias. When randomized controls are unavailable, however, valid causal inference requires additional assumptions relating the trial population to the external controls enabling adjustment of confounders. These assumptions, referred to as identifiability conditions, specify when a causal quantity of interest—such as $\tau_{\ATT}$ in \Eq{eq:att-def}—can be recovered from the observed data. Standard identifiability conditions for $\tau_{\ATT}$ are \cite{hernan2020, imbens2015, rubin1974}:

\begin{enumerate}
    \item \textbf{Consistency}. The observed outcome corresponds to the potential outcome under the treatment actually received:
    \begin{equation}
         Y = AY(1) + (1-A)Y(0).
    \end{equation}
    This assumption requires that treatment be well-defined (i.e., no ambiguity or multiple versions of treatment). Typically, one also requires no interference between participants, meaning one participant’s outcome does not depend on another participant’s treatment assignment. These conditions are often collectively referred to as the Stable Unit Treatment Value Assumption (SUTVA) and are typically ensured by study design.

    \item \textbf{Positivity}. Each treated participant has a non-zero probability of receiving control, conditional on baseline covariates:
    \begin{equation}
        P(A=0 \mid X=x) > 0 \quad \text{for all } x \text{ such that } P(X=x \mid A=1) > 0,
    \end{equation}
    where $X$ denotes baseline covariates. Positivity requires sufficient overlap between the covariate distributions of treated and control populations. While the distributions need not be identical, there must be adequate support in the control data for the covariate profiles observed among treated participants. This condition is essential for constructing valid comparators by adjusting for differences in baseline characteristics. For the ATT, this requirement is weaker than full positivity, as it only requires overlap for the treated population.  This means that when constructing controls we can consider, in principle, datasets broader than the study population and the particular adjustment methods discussed will ignore irrelevant contributions with no support on the study population.

    \item \textbf{Ignorability}. There are no unmeasured confounders; equivalently, conditional on observed covariates, treatment assignment is independent of potential outcomes:
    \begin{equation}
        (Y(0), Y(1)) \perp A \mid X.
    \end{equation}
    In randomized trials, this condition is guaranteed by design. In single-arm settings, it must be assumed, and its validity depends on the availability and quality of measured covariates. Given ignorability, appropriate adjustment for measured confounders is required to obtain unbiased estimates of treatment effects, while positivity ensures that such adjustment is feasible.  In practice, this may be the strongest assumption made, and we note that interpretable frameworks exist to quantify potential violations of ignorability~\cite{chernozhukov2024longstoryshortomitted}.
    
\end{enumerate}
Throughout most of this paper, we focus on $\tau_{\ATT}$ and assume identifiability. In \Secs{sec:strategy}{sec:digital-twins}, we briefly discuss cases in which model-based approaches can incorporate additional assumptions when identifiability is not guaranteed, although a comprehensive analysis is outside the scope of this work.

\section{Choosing an External Comparator Strategy}
\label{sec:strategy}
Before discussing common ATT estimators, it is helpful to review the main considerations in choosing external comparators. The choice of method depends on the availability, relevance, and quality of external data, as well as on the plausibility of identifiability assumptions. In this section, we outline a pragmatic framework for selecting among comparator strategies. Rather than a strict decision tree, the framework is best viewed as a sequence of considerations, where the answer at each stage informs the feasible set of approaches.

\begin{enumerate}

\item {\bf Is the comparator required to be tailored to the trial population?}
The first consideration is whether the external comparator must be informed by the characteristics of the enrolled population. By tailored we mean that the comparator is constructed to reflect
the specific covariate profile of the enrolled population, rather
than a generic reference value. A tailored comparator answers the
question \emph{what outcomes would these particular patients have
experienced under standard of care?} A non-tailored comparator instead uses a single
population-level summary (e.g.,\ a published mean survival time, a
clinically agreed reference value, or a fixed null change in a
biomarker) that does not depend on who is enrolled.

\begin{enumerate}
\item[$\circ$] If not, one may rely on fixed or externally defined benchmarks, such as estimates derived from network meta-analyses, historical summary statistics, or clinically informed reference values. These approaches require minimal assumptions about patient-level comparability, but do not adjust for differences in baseline characteristics and therefore may be limited in interpretability.
\begin{enumerate}
    \item[$-$] This may occur, for example, in an indication where there is a clinical consensus on average survival time and the treatment is believe to significantly lengthen survival.  Using the consensus survival time as the control outcome may be sufficient to show the treatment is effective, although precise treatment effects could not be obtained.
    \item[$-$] Another example is biomarkers in studies where the change in the value in the control arm is expected to be small.  A reference assumption of no change may be acceptable; however, one must be careful that variability in biomarker values can create effects such as regression to the mean which induce larger observed changes even if the underlying biological change is small.
\end{enumerate}

\item[$\circ$] If the comparator is required to reflect the enrolled population, patient-level adjustment becomes necessary, and methods based on external patient-level data should be considered.
\end{enumerate}

\item {\bf Is there external data that is plausibly exchangeable with a control population?}
The next consideration is whether external data can be regarded as exchangeable with the target control population, conditional on measured covariates. This corresponds to the plausibility of ignorability and overlap assumptions discussed in Section~\ref{sec:assumptions}.

\begin{enumerate}
\item[$\circ$] If exchangeability is not plausible, direct data-driven approaches, such as propensity score matching or weighting, are unlikely to yield valid estimates. This situation commonly arises when external data differ systematically from the trial setting. For example, observational data might lack placebo effects or real-world data used for external controls are subject to residual confounding relative to randomized trials. In such cases, model-based approaches that incorporate additional structural assumptions might become necessary. These methods can explicitly model corrections, for instance by recalibrating model outputs to population-level information from published trial data. While these approaches extend beyond what is directly supported by the observed patient-level data, they do so at the cost of increased model dependence. We plan to discuss these these extensions in a future publication.
\begin{itemize}
    \item[$-$] There may be many cases where structured assumptions become necessary because exchangeability is not credible, and these can overlap with cases where randomized experiments are not ethical or possible (e.g., due to challenges maintaining blinding).  It is helpful to view structured assumptions as a constructive path in these kinds of situations to be able to make quantitative statements, rather than something to be avoided.
    \item[$-$] Examples where these types assumptions apply is in an open-label setting where placebo effects from either the route of administration (such as surgery) or mechanism of action (such as adverse effects) create unblinding or the impossibility of a placebo arm, and different placebo responses are expected compared to more standard designs (e.g., an oral medication or injectible)~\cite{Benedetti2014-nn}.  Structured assumptions on the magnitude of placebo response can be layered on top of standard causal assumptions as a way to augment outcome models, and the assumptions tested with data from other studies or informed by clinical expertise.
    \item[$-$] Another example of a structured assumption is in rare disease, where parametric disease progression models can be used to describe both natural history progression and treatment effects~\cite{Wang2018-cf}. In many cases, these approaches allow for validation of assumptions from the observed trial data to determine their credibility \emph{post hoc}.
\end{itemize}

\item[$\circ$] If exchangeable external data are available, one can proceed to assess the quality and scope of that data.
\end{enumerate}

\item {\bf Is the external data sufficiently rich and of high quality?}
When external data are plausibly exchangeable, the choice of method depends on their quality, size, and coverage of the covariate space.

\begin{enumerate}
\item[$\circ$] If the data are limited in size, exhibit substantial missingness, or provide incomplete overlap with the trial population, relying solely on propensity-based adjustment may lead to unstable or biased estimates. In such cases, methods based on models that predict counterfactual outcomes, or combinations of these with propensity-based methods, such as through doubly robust estimators, can improve both robustness and efficiency. Outcome-model-based approaches can serve as primary or sensitivity analyses alongside purely propensity-based methods.
\begin{itemize}
    \item[$-$] This is the case this paper focuses on, and generally a case that many studies find themselves in.  The growing availability of relevant real-world, observational, and clinical trial data across indications means that outcome models can often be built, and as discussed there are advantages over pure data-based approaches alone.
\end{itemize}

\item[$\circ$] If the external data are sufficiently large, high-quality, and well-aligned with the trial population, propensity-based methods alone may be adequate to construct a valid comparator, provided that the propensity model is appropriately specified and overlap conditions are satisfied. Outcome-model-based methods can still provide efficiency gains.
\begin{itemize}
    \item[$-$] We note that applications need not choose a single approach.  For example, even if a trialists arrive at this last question, both propensity-based and model-based methods may be worthwhile to consider as they provide different advantages and lines of evidence, and their agreement can be used as a way to at least partially validate each other.
\end{itemize}
\end{enumerate}
\end{enumerate}

We might encapsulate the last question with the adage \emph{use data when you can, models when you must}.  Data is often a tempting comparator due to the fact that it is tangible.  However, one must be careful that data is exchangeable and is well understood, so that the causal assumptions hold for propensity-based estimators.  Model-based approaches, particularly those based on flexible outcome models, play a central role when data are heterogeneous, limited, or only partially aligned with the target population, as discussed above. 

\begin{figure}[t!]
\centering
\includegraphics[width=0.8\linewidth]{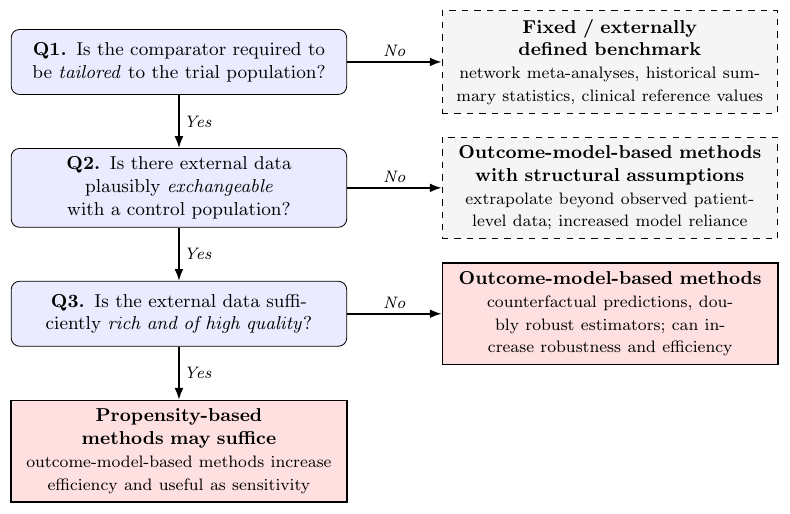}
\caption{Decision flow for selecting an external comparator strategy
in single-arm trials, summarizing \Sec{sec:strategy}. The two red boxes correspond to
settings where patient-level external data permit direct estimation;
outcome-model-based estimators are particularly attractive when external data are limited or heterogeneous, and can provide complementary evidence alongside propensity-based methods.}
\label{fig:comparator-strategy}
\end{figure}

In \Fig{fig:comparator-strategy}, we summarize the workflow described in this section. In practice, conclusions at one stage may prompt reconsideration of earlier choices. For example, if no exchangeable external data are available, one may revert to fixed or qualitative comparators. Similarly, if available data are insufficient to support reliable propensity-based adjustment, model-based approaches may become the primary strategy even when exchangeability is plausible. Overall, this framework highlights that the choice of comparator is driven by a trade-off between reliance on data and reliance on modeling assumptions.  We will return to the influence of outcome models when discussing the FDA framework for model development, risk, and credibility in \Sec{sec:fda-guidance}. 

\section{Treatment Effect Estimators}
\label{sec:estimators}

In this section, we review the definitions and properties of commonly
used ATT estimators in single-arm trials. Our goal is not to provide a
comprehensive survey of available estimators, but
rather to characterize a few well-understood classes and compare them
in terms of the requirements
each class places on its underlying nuisance models for consistent
estimation and the form of its
asymptotic variance under simplifying assumptions. Together these
determine when an estimator can be trusted and how efficient it is
expected to be at a given sample size.

We organize the discussion around two complementary classes:
propensity-based estimators in \Sec{sec:propensity}, which adjust for confounding by matching
or reweighting on an estimated treatment-assignment model, and
outcome-model-based estimators in \Sec{sec:plugin}, which model the conditional response
under control and combine the resulting predictions with the trial
data either directly or with a propensity-based correction.  
Within each of these
classes there are many variants (e.g., different matching schemes, different normalizations, different forms of the doubly robust estimator), but
to keep the comparison
clear we focus on a simple representative of each class:
one-to-one propensity score matching, inverse probability weighting,
the plug-in estimator from a fitted outcome model, and the augmented
inverse probability weighting estimator.

\subsection{Propensity-Based Estimators}
\label{sec:propensity}

A common approach to estimating treatment effects using external
data is based on propensity scores. The propensity score is defined
as the probability of treatment assignment conditional on baseline
covariates,
\begin{equation}
    e(X) = P(A=1 \mid X),
\end{equation}
where $A \in \{0,1\}$ denotes treatment assignment and $X$ represents
observed baseline covariates. Intuitively, $e(X)$ is the probability
that a patient with covariate profile $X$ is observed in the treated
group rather than in the external control group. By construction,
$e(X)$ summarizes in a single scalar the measured covariate
information that predicts treatment assignment. In a single-arm
trial, $e(X)$ is unknown and is typically estimated from the pooled
dataset of treated and external control subjects, by fitting a binary
classifier of $A$ on $X$. Under ignorability and positivity, the
propensity score is a balancing score: conditional on $e(X)$, the
distribution of covariates is the same between treated and control
groups \cite{rosenbaum1983}. This enables adjustment for confounding
through weighting or matching.

Propensity models are useful because they enable comparisons between two overlapping populations by accounting for differences in their composition within the region of overlap. For example, if two studies use the same basic inclusion criteria but enroll different patient populations, a propensity model can account for those shifts and support comparisons of outcomes.

\paragraph{Propensity score matching.}
One approach to estimate ATT in a single-arm trial is to match treated participants to control participants
with similar propensity scores. Let $m(i)$ denote the set of matched
control indices for treated participant $i$, and let $|m(i)|$ denote
its size. A matching estimator of the ATT can be written as
\begin{equation}
\label{eq:psm}
    \hat{\tau}_{\ATT}^{\text{PSM}} = \frac{1}{n_1} \sum_{i: A_i=1}
    \left( Y_i - \frac{1}{|m(i)|} \sum_{j \in m(i)} Y_j \right),
\end{equation}
where the inner average represents the matched control outcome for
treated participant $i$.

In the simplest case of one-to-one matching without replacement,
$|m(i)| = 1$, and we recover a standard nearest-neighbor estimator.
More generally, matching schemes may include multiple controls per
treated participant and/or weighting within matches.

Even with a well-specified propensity model, matching estimators can
suffer from bias due to imperfect matches. In particular,
fixed-$M$ nearest-neighbor matching is generally biased unless
additional smoothness conditions hold or a bias correction is applied.
A detailed discussion about matching estimators can be found, e.g., in \cite{abadie2006, stuart2010matching}. As a reference for the asymptotic variance, consider
the case of one-to-one matching without replacement. Assuming a
well-specified propensity model and constant treatment effect, the
asymptotic variance of the estimator is
\begin{equation}
    \V[\widehat{\tau}_{\ATT}^{\text{PSM}}] \approx \frac{2\sigma_1^2}{n_1},
    \,\text{or}\;\;
    \V[\widehat{\tau}_{\ATT}^{\text{PSM}}] \approx \frac{2\kappa^2}{n_1},
\end{equation}
where $\sigma_1^2$ is the marginal variance of the outcome in the
treated group which we have assumed to be roughly equivalent to the
marginal variance of the matched control, with a constant treatment
effect. If an outcome model is used to remove matching bias, and an
equal homoskedastic error is assumed in the treated and control, the
variance can be written in terms of the common conditional variance
$\kappa^2=\kappa_0^2=\kappa_1^2=\V[Y\mid X, A=0] = \V[Y\mid X, A=1]$.

\paragraph{Inverse probability weighting.}
An alternative approach uses inverse probability weighting (IPW) to
reweight control observations so that their covariate distribution
matches that of the treated group. For the ATT, the IPW estimator
can be written as
\begin{equation}
\label{eq:ipw}
    \hat{\tau}_{\ATT}^{\text{IPW}} = \frac{1}{n_1} \sum_{i: A_i=1} Y_i
    - \frac{1}{n_1} \sum_{i: A_i=0} \frac{e(X_i)}{1 - e(X_i)} Y_i,
\end{equation}
where $n_1 = \sum_i \mathbb{1}(A_i=1)$ is the number of treated
participants and $e(X_i)$ are the propensity scores. The propensity
odds $w(X) = e(X)/(1-e(X))$ are proportional to the density ratio
between the treated and control covariate distributions, so the
second term in \Eq{eq:ipw} can be interpreted as an importance-sampling
estimator of the average control outcome on the treated population;
the derivation is given in App.~\ref{app:ipw}.
If the propensity score $e(X)$ is known, the IPW estimator is
asymptotically normal,
\begin{equation}
\label{eq:ipw-asympt-normal}
    \sqrt{n}\!\left(\hat{\tau}_{\ATT}^{\text{IPW}} - \tau_{\ATT}\right)
    \xrightarrow{\;d\;}
    \mathcal{N}\!\left(0,\,\sigma^2_{\text{IPW}}\right),
\end{equation}
where $n=n_1+n_0$ and $\sigma^2_{\text{IPW}}$ is the asymptotic variance of the scaled estimator. Assuming equal homoskedastic errors in
the treated and control populations,
$\kappa^2 = \V[Y\mid X, A=0] = \V[Y\mid X, A=1]$, the variance of the
estimator simplifies to
\begin{equation}
\label{eq:ipw-var-simplified}
    \V\!\left[\widehat{\tau}_{\ATT}^{\text{IPW}}\right]
    \;\approx\;
    \frac{\sigma_1^2}{n_1} \;+\; \frac{\kappa^2 + \Delta}{n_0^{\mathrm{eff}}},
\end{equation}
where $\sigma_1^2$ is the marginal outcome variance in the treated
population. The effective control sample size $n_0^{\mathrm{eff}}$
accounts for loss of efficiency due to reweighting, and is given by
\begin{align}
\label{eq:neff}
n_0^{\mathrm{eff}} &\;=\; \gamma\,n_0,\\
\label{eq:gamma}
\gamma &\;=\; \left(\frac{p_1}{1-p_1}\right)^{2}\frac{1}{
\E\!\left[w^2(X) \mid A=0\right]}=\frac{1}{1+\chi^2\left(P_1(X)||P_0(X)\right)}
       \;\leq\; 1,
\end{align}
where $p_1 = P(A=1)$ estimated by the fraction of treated participants. In the last equality, we have used the $\chi^2$ divergence where $P_{0, 1}(X)=P(X\mid A=0,1)$ are the covariate distributions in the treated and control populations. The correction term $\Delta$ accounts for residual heterogeneity in
$\mu_0(X)$,
\begin{equation}
\label{eq:delta}
    \Delta = \frac{\V[w(X)\mu_0(X)\mid A=0]}{\E[w^2(X)\mid A=0]},
    \qquad \mu_0(X)=\E[Y\mid X, A=0].
\end{equation}
The derivation of the asymptotic properties of the IPW estimator is reported in App.~\ref{app:ipw}. In a single-arm trial setting, the propensity score is unknown and
must be estimated. The IPW estimator remains consistent for
$\tau_{\ATT}$ if the propensity score model $\hat{e}(X)$ is correctly
specified, in the sense that
\begin{equation}
    \|\hat{e} - e\|
    = \left( \E[(\hat{e}(X) - e(X))^2] \right)^{1/2}
    \xrightarrow{\;p\;} 0.
\end{equation}
However, consistency alone is not sufficient for valid statistical
inference. For asymptotic normality and for confidence intervals to
have correct coverage, the estimation error should be negligible
relative to sampling variability:
\begin{equation}
    \|\hat{e} - e\| = o_p(n^{-1/2}).
\end{equation}
This is typically a strong requirement on the model. Alternatively, if the
propensity model is parametric and well-specified, the variance can be computed using
M-estimation with stacked estimating equations ~\cite{hirano2003efficient}, and will account for
propensity estimation errors of order $O_p(n^{-1/2})$. Bootstrap methods which
include refitting the propensity model within each resample can also be
used to approximate the sampling distribution. The validity of these
methods still relies on the propensity model being well-specified with bias negligible at the
$n^{-1/2}$ scale.

Both matching and weighting rely on accurate estimation of the
propensity score. In practice, model misspecification or finite-sample
estimation error can lead to biased estimates. These limitations motivate the
use of complementary approaches, such as outcome modeling and doubly
robust estimators, which we discuss next.

\subsection{Outcome Model-Based Estimators}
\paragraph{Plug-in Estimator}
\label{sec:plugin}
An alternative to propensity-based approaches is to directly model
the outcome under control. Let
\[
\mu_0(X) = \E[Y \mid X, A=0]
\]
denote the conditional mean outcome under control. An outcome-model plug-in estimator of the ATT is given by
\begin{equation}
\label{eq:plugin}
    \hat{\tau}_{\ATT}^{\text{OM}}
    = \frac{1}{n_1} \sum_{i: A_i=1}
    \left( Y_i - \mu_0(X_i) \right).
\end{equation}
This estimator replaces the unobserved counterfactual outcomes
$Y_i(0)$ with predictions from the outcome model. With a known outcome model, the
estimator is asymptotically normal,
\begin{equation}
\label{eq:plugin-asympt-normal}
    \sqrt{n_1}\!\left(\hat{\tau}_{\ATT}^{\text{OM}} - \tau_{\ATT}\right)
    \xrightarrow{\;d\;}
    \mathcal{N}\!\left(0,\,\sigma^2_{\text{OM}}\right),
\end{equation}
where $\sigma^2_{\mathrm{OM}}$ is the asymptotic variance of the scaled estimator. Assuming a constant treatment effect and equal homoskedastic error in the treated and control, the
variance of the estimtor simplifies to
\begin{equation}
\label{eq:plugin-var-simplified}
    \V[\hat{\tau}_{\ATT}^{\text{OM}}]
    \;\approx\;
    \frac{\kappa^2}{n_1},
\end{equation}
where $\kappa^2 = \kappa_0^2 = \kappa_1^2 = \V[Y\mid X, A=0]
= \V[Y\mid X, A=1] = \sigma_0^2(1-\rho_0^2)$ is the
conditional variance of the outcome assumed to be constant under
homoskedasticity. It can also be rewritten in terms of the marginal
variance $\sigma_0^2$ and the correlation
$\rho_0=\mathrm{Corr}[\mu_0(x), Y\mid A=0]$. 

In practice, $\mu_0(X)$ must be estimated with a model
$\hat{\mu}_0(X)$. The outcome-model plug-in estimator for ATT remains consistent if
\begin{equation}
    \|\hat{\mu}_0 - \mu_0\|
    = \left( \E[(\hat{\mu}_0(X) - \mu_0(X))^2] \right)^{1/2}
    \xrightarrow{\;p\;} 0.
\end{equation}
However, as for the propensity model case, consistency alone is not
enough for valid inference. Modeling error must be negligible
compared to the scale of fluctuations to rely on asymptotic normality
with the variance derived above:
\begin{equation}
    \|\hat{\mu}_0 - \mu_0\| = o_p(n_1^{-1/2}).
\end{equation}
This is can be difficult to achieve or ensure. As we discuss in the
next section, doubly robust estimators relax these
requirements and enable valid inference under weaker conditions.

\paragraph{Doubly Robust Estimators}
\label{sec:aipw}

Doubly robust estimators combine outcome modeling with propensity
score weighting to improve robustness and enable valid inference
under weaker conditions. A canonical example is the augmented
inverse probability weighting (AIPW) estimator, which can be derived
from the efficient influence function for the ATT. In
App.~\ref{app:eif}, we derive the efficient influence function and
obtain the corresponding estimator:
\begin{equation}
\label{eq:aipw}
\hat{\tau}_{\ATT}^{\text{AIPW}}
=
\frac{1}{n_1} \sum_{i: A_i=1} \left( Y_i - \mu_0(X_i) \right)
-
\frac{1}{n_1} \sum_{i: A_i=0} \frac{e(X_i)}{1 -e(X_i)}
\left( Y_i - \mu_0(X_i) \right).
\end{equation}
The first term corresponds to the outcome-model plug-in estimator, while the second
term corrects for bias in the outcome model using reweighted
residuals from the control group. As in IPW, the propensity odds
$w(X) = e(X)/(1-e(X))$ are proportional to the density ratio between
treated and control covariate distributions and can be interpreted as
importance weights. Here, they are used to reweight residuals
computed under the control distribution $P(X \mid A=0)$ to
approximate their expectation under the treated distribution
$P(X \mid A=1)$.

We refer to, e.g., \cite{tsiatis2006semiparametric, chernozhukov2018, kennedy2022}, for a
detailed discussion of AIPW and doubly robust estimators. For the
purposes of this work, we would like to highlight the following key
properties:

\begin{enumerate}

\item \textbf{Double robustness.}
In any single-arm trial setting, conditional means and propensity
scores need to be estimated. One advantage of this estimator is that
it remains consistent for $\tau_{\ATT}$ if either the outcome model
$\hat{\mu}_0(X)$ or the propensity score model $\hat{e}(X)$ is
correctly specified, but not necessarily both, providing additional
protection against misspecification:
\[
\|\hat{\mu}_0 - \mu_0\| \xrightarrow{\;p\;} 0
\quad \text{or} \quad
\|\hat{e} - e\| \xrightarrow{\;p\;} 0.
\]

\item \textbf{Second-order bias and weaker rate conditions.}
When both models are correctly specified, the leading-order bias is
removed by construction, and the estimator depends on nuisance
estimation error only at second order. Specifically, asymptotic
normality holds if
\begin{equation}
\label{eq:aipw-convergence}
\|\hat{\mu}_0 - \mu_0\| \cdot \|\hat{e} - e\| = o_p(n^{-1/2}),
\end{equation}
which is satisfied, for example, if both nuisance estimators converge
at rate $o_p(n^{-1/4})$, which is more feasible when using machine
learning models. Under these conditions, the AIPW estimator is
asymptotically normal,
\begin{equation}
\label{eq:aipw-asympt-normal}
    \sqrt{n}\!\left(\hat{\tau}_{\ATT}^{\text{AIPW}} - \tau_{\ATT}\right)
    \xrightarrow{\;d\;}
    \mathcal{N}\!\left(0,\,\sigma^2_{\text{AIPW}}\right),
\end{equation}
where $\sigma^2_{\text{AIPW}}$ is the variance of the efficient
influence function derived in App.~\ref{app:eif}. Assuming a constant
treatment effect and the same homoskedastic error across arms, we
show in App.~\ref{app:variance} that the variance simplifies to
\begin{equation}
\label{eq:att_hh}
    \V\!\left[\widehat{\tau}_{\ATT}^{\text{AIPW}}\right]
    \;\approx\;
    \kappa^2\!\left(\frac{1}{n_1}+\frac{1}{n_0^{\mathrm{eff}}}\right),
\end{equation}
where $\kappa^2 = \kappa_0^2 = \kappa_1^2 = \V[Y\mid X, A=0]
= \V[Y\mid X, A=1]$ is the conditional variance of the outcome
assumed to be constant under homoskedasticity and equal in the
treated and control populations. The conditional variance can also
be rewritten, e.g., as
$\kappa_0^2 = \V[Y\mid X, A=0] = \sigma_0^2(1-\rho_0^2)$ in terms of
the marginal variance $\sigma_0^2$ and the correlation
$\rho_0=\mathrm{Corr}[\mu_0(x), Y\mid A=0]$, assuming a constant
treatment effect. The effective control sample size
$n_0^{\mathrm{eff}}=\gamma n_0$ is defined in
\Eq{eq:gamma}. The factor $\gamma$ captures the loss of
efficiency due to differences in covariate distributions between
treated and control populations. When the distributions are the same
($\gamma = 1$), we recover the variance of a randomized controlled
trial with outcome model adjustment, as in PROCOVA
\cite{schuler2022prognostic}.

\item \textbf{Semiparametric efficiency.}
If both nuisance models are correctly specified and satisfy the
convergence conditions described above, the AIPW estimator achieves
the semiparametric efficiency bound, meaning that it has the smallest
possible asymptotic variance among all regular and asymptotically
linear estimators of $\tau_{\ATT}$. This optimality follows from the
fact that the estimator is based on the efficient influence function.
\end{enumerate}
AIPW is not the only doubly robust estimator. Another widely used
class is based on targeted maximum likelihood estimation (TMLE)
\cite{vanderlaan2006, vanderlaan2011}. TMLE starts from an initial
estimate of the outcome model and updates it through a targeted
fitting step that incorporates the propensity score, ensuring that
the resulting estimator solves the efficient influence function
estimating equation. Like AIPW, TMLE is doubly robust and achieves
semiparametric efficiency under standard conditions, and, in some
cases, can have better finite-sample properties. While we focus on
AIPW due to its simplicity and direct interpretability, TMLE provides
an alternative framework with similar theoretical guarantees.

In Tab.\,\ref{tab:estimators} we summarize the key properties of the
ATT estimators discussed in this section.

\begin{table}[t]
\centering
\small
\renewcommand{\arraystretch}{1.2}
\caption{Summary of estimators for the ATT discussed in
\Sec{sec:estimators}. For the asymptotic variance, we assume constant
treatment effect and equal homoskedastic error in the treated and
control groups. $\sigma_{1}^2$ is the marginal variance of the
outcome in the treated group, assuming a
constant treatment effect. $\kappa^2$ is the conditional variance
assumed to be constant under homoskedasticity and equal across arms,
$\kappa^2=\kappa_0^2=\kappa_1^2=\V[Y\mid X,A=0]=\V[Y\mid X, A=1]$.
$n_1$ is the treated sample size and $n_0^{\mathrm{eff}}=\gamma n_0$
is the effective control sample size, defined in
\Eqs{eq:neff}{eq:gamma}. $\Delta$ is defined in \Eq{eq:delta}. For
propensity score matching we have assumed one-to-one matching without
replacement, for simplicity. If nuisance models (propensity or outcome) are parametric and well-specified, then estimation errors of order $O_p(n^{-1/2})$ can be taken into account via M-estimation.}
\begin{tabular}{l p{4cm} p{4.5cm} p{3.5cm}}
\toprule
Estimator
& Consistency
& Inference requirement
& Asymptotic variance\\
\midrule

PSM
& $\hat{e}(X)$ consistent + matching bias correction
& $\|\hat e - e\| = o_p(n^{-1/2})$ + matching bias = $o_p(n^{-1/2})$
& $\frac{2\sigma_{1}^2}{n_1}$ or $\frac{2\kappa^2}{n_1}$ \\

IPW
& $\hat{e}(X)$ consistent
& $\|\hat e - e\| = o_p(n^{-1/2})$
& $\frac{\sigma_{1}^2}{n_1}+\frac{\kappa^2+\Delta}{n_0^{\mathrm{eff}}}$ \\

OM
& $\hat{\mu}_0(X)$ consistent
& $\|\hat\mu_0 - \mu_0\| = o_p(n_1^{-1/2})$
& $\frac{\kappa^2}{n_1}$ \\

AIPW
& $\hat{e}(X)$ or $\hat{\mu}_0(X)$ consistent
& $\|\hat e - e\|\cdot\|\hat\mu_0 - \mu_0\| = o_p(n^{-1/2})$
& $\kappa^2\!\left(\frac{1}{n_1} + \frac{1}{n_0^{\mathrm{eff}}}\right)$ \\

\bottomrule
\end{tabular}
\label{tab:estimators}
\end{table}

\section{Why Digital Twins?}
\label{sec:digital-twins}
In this section, we discuss when and why outcome-model–based approaches—particularly those based on digital twins—are preferable to propensity-based methods. A \emph{digital twin} is a patient-level model that provides a comprehensive prediction of future outcomes over time under a specified standard of care, conditional on baseline characteristics. In other words, for each patient, the digital twin represents a personalized estimate of their counterfactual trajectory in the absence of treatment across a number of outcomes of interest. In this work, we focus on digital twins generated from machine learning models trained on historical data.  

Having introduced several estimators for single-arm trials, we now compare their properties and highlight settings in which outcome modeling provides practical and theoretical advantages.  At a high level, both propensity-based methods and digital twin approaches aim to adjust for measured confounders so that, in the absence of unmeasured confounding, differences in outcomes between treated patients and external controls can be attributed to the treatment. The key distinction lies in how this adjustment is performed: propensity-based methods reweight or match observations to balance covariates, whereas digital twin models directly estimate counterfactual outcomes conditional on covariates. Below we summarize key advantages of model-based approaches and their implications for treatment effect estimation.

\paragraph{Advantages of digital twin approaches.}
Outcome-model–based approaches offer several advantages, particularly when combined with flexible machine learning methods:

\begin{enumerate}

\item \textbf{Flexibility to leverage heterogeneous data.}
Modern machine learning models can integrate data that would be difficult to use directly in propensity-based analyses. This includes heterogeneous datasets with differing covariate sets, varying outcome measurement cadence, and complex missingness patterns, as well as multi-source data (e.g., clinical trials and real-world data), multi-resolution data (e.g., patient-level and aggregate summaries), and multimodal inputs (e.g., structured variables, images, and text). 

Digital twin models can learn shared structure across these sources and borrow strength across datasets, allowing partial information from each source to contribute to the final model. In contrast, propensity-based methods typically require aligned covariates and comparable data structures, limiting the usable data. As a result, outcome models can make more efficient use of available historical information and reduce reliance on narrowly defined subsets of data.

\item \textbf{Ability to leverage larger and more diverse datasets.}
Propensity-based methods generally require restricting attention to data whose covariate distributions overlap with the target population to satisfy positivity, often discarding substantial portions of available data. In contrast, outcome models can be trained on broader datasets spanning multiple populations. 

In particular, modern machine learning approaches enable forms of transfer learning, where models trained on large and diverse datasets learn generalizable representations that can be fine-tuned and adapted to the target population. This allows digital twin models to leverage information from related populations even when they are not perfectly aligned with the target cohort. Such approaches improve predictive performance and reduce the risk of misspecification, while still allowing conditioning on the characteristics of the trial population at inference time.

\item \textbf{Application to multiple endpoints.}
In many studies, endpoints beyond the primary endpoint are of interest for assessing efficacy or understanding disease course. Digital twin models can typically provide comprehensive predictions across multiple endpoints by leveraging broader training data. This is an advantage over propensity-based methods, especially when data overlapping with the target trial population are limited or support only a small subset of endpoints.

\item \textbf{Natural validation framework.}
Outcome models performance can be evaluated directly against observed outcomes in the historical data using standard predictive metrics (e.g., bias and mean squared error) on held-out or cross-validated samples. This provides a transparent and quantitative way to verify that the model captures relevant outcome relationships.

In contrast, propensity score models predict the treatment assignment mechanism, which is not directly observable in observational data. As a result, they are typically evaluated indirectly through balance diagnostics.

\end{enumerate}

\paragraph{Implications for treatment effect estimation.}
We now discuss how these advantages translate into improved properties of treatment effect estimators.

\begin{enumerate}

\item \textbf{Increased efficiency.}
Both outcome-model plug-in and AIPW estimators predict counterfactual outcomes for each treated participant. This induces a paired structure, where each observed outcome is compared to its predicted counterfactual. As a result, variability is reduced relative to estimators based on matching or weighting raw outcomes. 

This reduction in variance is reflected in the conditional variance appearing in the asymptotic formulas in Tab.~\ref{tab:estimators}. If we assume the same homoskedastic error in the treated and control populations and a constant treatment effect, then the conditional variance can be estimated, e.g., from the control population as $\kappa^2=\kappa_0^2=\V[Y\mid X, A=0] = \sigma_0^2(1-\rho_0^2)$ in terms of the marginal variance $\sigma_0^2$ and the correlation $\rho_0=\mathrm{Corr}[\mu_0(x), Y\mid A=0]$ of the outcome model with observed outcomes. When both nuisance functions are correctly specified and satify the convergence requirements in \Eq{eq:aipw-convergence}, estimators such as AIPW achieve the semiparametric efficiency bound.

\item \textbf{Improved robustness to misspecification.}
Propensity-based estimators are consistent when the propensity score model is correctly specified. Similarly, outcome-model plug-in estimators rely on correct specification of the outcome model. 

Outcome models trained on large and diverse datasets, particularly with flexible machine learning methods, are likely less prone to misspecification. Moreover, outcome modeling enables the use of doubly robust estimators such as AIPW, which remain consistent if either the outcome model or the propensity model is correctly specified. This provides protection against misspecification of either component and improves reliability.

\item \textbf{Weaker convergence conditions for valid inference with doubly robust estimators.}
While consistency requires correct specification of nuisance functions, valid statistical inference requires stronger conditions on their rates of convergence.

Propensity-based and outcome-model plug-in estimators typically require estimation errors to satisfy
\[
\|\hat{\mu}_0 - \mu_0\| = o_p(n_1^{-1/2}) \quad \text{or} \quad \|\hat{e} - e\| = o_p(n^{-1/2}),
\]
where $n$ is the total sample size. These are strong conditions and generally require either fast convergence rates or very large training datasets. For example, if an outcome model converges at rate $O_p(n_{\text{train}}^{-1/4})$, achieving $o_p(n^{-1/2})$ would require $n_{\text{train}} \gg n^2$, which is often unrealistic.

Doubly robust estimators such as AIPW relax this requirement by requiring only that
\[
\|\hat{\mu}_0 - \mu_0\| \cdot \|\hat{e} - e\| = o_p(n^{-1/2}).
\]
This weaker condition allows the use of flexible machine learning models while still obtaining asymptotically normal estimators and valid confidence intervals. In particular, under these condtitions
\[
\sqrt{n}(\hat{\tau}_{\ATT}^{\text{AIPW}} - \tau_{\ATT}) \xrightarrow{d} \mathcal{N}(0, \sigma_{\text{AIPW}}^2),
\]
and 95\% confidence intervals can be constructed as $\hat{\tau}_{\ATT}^{\text{AIPW}}\pm 1.96\sqrt{\hat{\sigma}^2_{\mathrm{AIPW}}/n}$, where $\hat{\sigma}^2_{\mathrm{AIPW}}$ is the estimated variance of the efficient influence function.

\item \textbf{Extensions beyond direct identifiability.}
In settings where patient-level external data are available but not directly comparable to the trial population, outcome-model–based approaches provide a flexible framework for incorporating additional assumptions. For example, as mentioned in \Sec{sec:strategy}, observational and real-world data used for external controls might lack placebo effects or might be subject to
residual confounding relative to randomized trials. 
Outcome models can explicitly attempt to remove discrepancies, or learn corrections, for instance by recalibrating to summary statistics from published trial data.  Other examples are also discussed in \Sec{sec:strategy}.

While such approaches rely on stronger assumptions (e.g., transportability of the calibration), they enable estimation in settings where direct comparisons are not feasible. Importantly, they also support sensitivity analyses, allowing practitioners to assess the robustness of conclusions to these assumptions. While these applications are beyond the scope of this manuscript, they will be discussed in future work.
\end{enumerate}

\section{Power and Sample Size for Outcome-Model Estimators}
\label{sec:power}

In this section, we present power and sample size formulas for estimators that use outcome models, as well as a focused discussion of the role of additional historical data. We note that adding more data should, in principle, never increase the variance of the treatment effect estimate, but its potential benefits must be weighed against added cost and complexity.

Power and sample size computations for outcome-model-based approaches follow from
the asymptotic distribution of the underlying estimator. We focus on
AIPW as a natural setup. It is also conservative relative
to the plug-in estimator, which achieves smaller asymptotic variance
only under the stronger assumption that the outcome model is correctly
specified with negligible errors. Assuming a constant treatment effect and the same homoskedastic error in the treated and external control groups, we showed in App.~\ref{app:variance} that the asymptotic variance of the AIPW estimator is given by:
\begin{equation}
\V[\widehat{\tau}_{ATT}^{\text{AIPW}}] \;\approx\;
\kappa^2\left(\frac{1}{n_1}+\frac{1}{n_0^{\mathrm{eff}}}\right),
\label{eq:aipw-variance-power}
\end{equation}
where $\kappa^2=\kappa_0^2=\kappa_1^2=\V[Y\mid X, A=0] = \V[Y\mid X, A=1]$ is the conditional variance of the outcome assumed to be constant under homoskedasticity and equal in the treated and control populations. The conditional variance can also be rewritten, e.g., as $\kappa^2 = \V[Y\mid X, A=0] = \sigma_0^2(1-\rho_0^2)$ in terms of the marginal variance $\sigma_0^2$ and the correlation $\rho_0=\mathrm{Corr}[\mu_0(x), Y\mid A=0]$, assuming a constant treatment effect. Both
quantities can be estimated from an external control population with
similar inclusion criteria as the target trial: $\sigma_{0}^2$
directly from the control outcomes, and $\rho_0$ from the out-of-sample correlation between the fitted outcome model and the observed responses. The effective control sample size $n_0^{\mathrm{eff}}=\gamma n_0$ is defined in \Eq{eq:neff}.
Using the asymptotic normality of the AIPW estimator, power for a
two-sided test with significance $\alpha$, and assuming a true effect $\tau$, follows the
standard continuous-outcome formula
\begin{equation}
\text{Power} \;=\; \Phi\!\left(\Phi^{-1}(\alpha/2)+\frac{\tau}{\sqrt{\V[\widehat{\tau}_{ATT}^{\text{AIPW}}]}}\right) + \Phi\!\left(\Phi^{-1}(\alpha/2)-\frac{\tau}{\sqrt{\V[\widehat{\tau}_{ATT}^{\text{AIPW}}]}}\right),
\label{eq:power-formula}
\end{equation}
where $\Phi$ is the CDF of the standard normal distribution. Computation of sample size $n_1$ proceeds as for a standard RCT given
estimates of $\rho_0$, $\kappa$, $n_0$, and $\gamma$. In what follows
we take the conservative case $\rho_0 = 0$, so $\kappa^2$ reduces to
the marginal outcome variance $\sigma_{0}^2$ on a control cohort with a population similar to the target trial.

\paragraph{Estimating $n_0^{\mathrm{eff}}$ prospectively.}
The main remaining
quantity to estimate prospectively is $\gamma$, which represents the effective fraction of the external control data, after adjusting for baseline distribution differences. It can 
can be written as
\begin{equation}
\gamma \;=\; \left(\frac{p_1}{1-p_1}\right)^{2}\frac{1}{
\E\!\left[w^2(X) \mid A=0\right]}=\frac{1}{1+\chi^2\left(P_1(X)||P_0(X)\right)},
\label{eq:lambda-alt}
\end{equation}
where $P_{0, 1}(X)=P(X\mid A=0,1)$ are the covariate distributions in the treated and control populations. Estimating $\gamma$ requires a working model for the covariate
mismatch. When patient-level data that approximate the planned trial
cohort are available---for example, data from a prior trial in the same
indication with similar inclusion criteria ---one can fit a propensity model $\hat{e}(X)$ comparing the proxy trial baseline distribution with the historical data that meet the inclusion criteria and
estimate
\begin{equation}
\hat\gamma \;=\; \frac{n_1^2}{n_0}\left(
\sum_{i:A_i=0}\!\! \hat w^2(X_i)\right)^{-1},
\qquad \hat w(X) = \frac{\hat e(X)}{1-\hat e(X)}.
\label{eq:lambda-pilot}
\end{equation}
When no such data are available, a possible heuristic is to approximate
the trial and external populations as Gaussian with comparable
covariance on the key prognostic covariates. Under this approximation,
$\chi^2\!\big(P_1(X)||P_0(X)\big) =
\exp(d_M^2) - 1$, and hence
\begin{equation}
\gamma \;\approx\; \exp\!\big(-d_M^2\big),
\qquad
d_M^2 \;=\; (\mu_T - \mu_H)^{\!\top}\, \Sigma^{-1}\, (\mu_T - \mu_H),
\label{eq:lambda-smd}
\end{equation}
where $d_M$ is the Mahalanobis distance between the covariate means of
the two populations. If the covariates are further assumed
to be approximately uncorrelated, Eq.~\eqref{eq:lambda-smd} reduces to
a sum of squared standardized mean differences (SMDs),
$d_M^2 \approx \sum_j \mathrm{SMD}_j^2$. As a rough calibration, SMDs below $0.1$ are typically considered
well-balanced, between $0.1$ and $0.25$ moderately imbalanced, and
above $0.25$ strongly imbalanced. As a reference, under this approximation, ten moderately
imbalanced covariates yield $\gamma \approx 1/2$; in practice, in the amyotrophic lateral sclerosis and Huntington's disease cases studies analyzed in \Sec{sec:case-studies}, we estimate $\gamma$ to be approximately in the range $0.2 - 0.6$, depending on the endpoint analyzed. For a binary covariate with trial prevalence $\pi_1$ and historical
prevalence $\pi_0$, the $\chi^2$ contribution admits a closed form and under approximate
independence across covariates, continuous and binary covariates can be combined by summing their contributions to $d_M^2$:
\begin{equation}
\gamma \;\approx\; \exp\!\left(-\sum_{j\in\text{cont.}} \mathrm{SMD}_j^2
\;-\; \sum_{j\in\text{bin.}} \frac{(\pi_{1,j}-\pi_{0,j})^2}{\pi_{1,j}(1-\pi_{1,j})}\right).
\label{eq:lambda-mixed}
\end{equation}
Because the true $\gamma$ is more sensitive to the density-ratio tails
than the Gaussian approximation, we recommend using
Eq.~\eqref{eq:lambda-mixed} as a heuristic only, and
preferring the propensity estimate of Eq.~\eqref{eq:lambda-pilot} whenever feasible. In both cases, all known confounders should be included.

\begin{figure}[t!]
    \centering
    \includegraphics[width=0.95\linewidth]{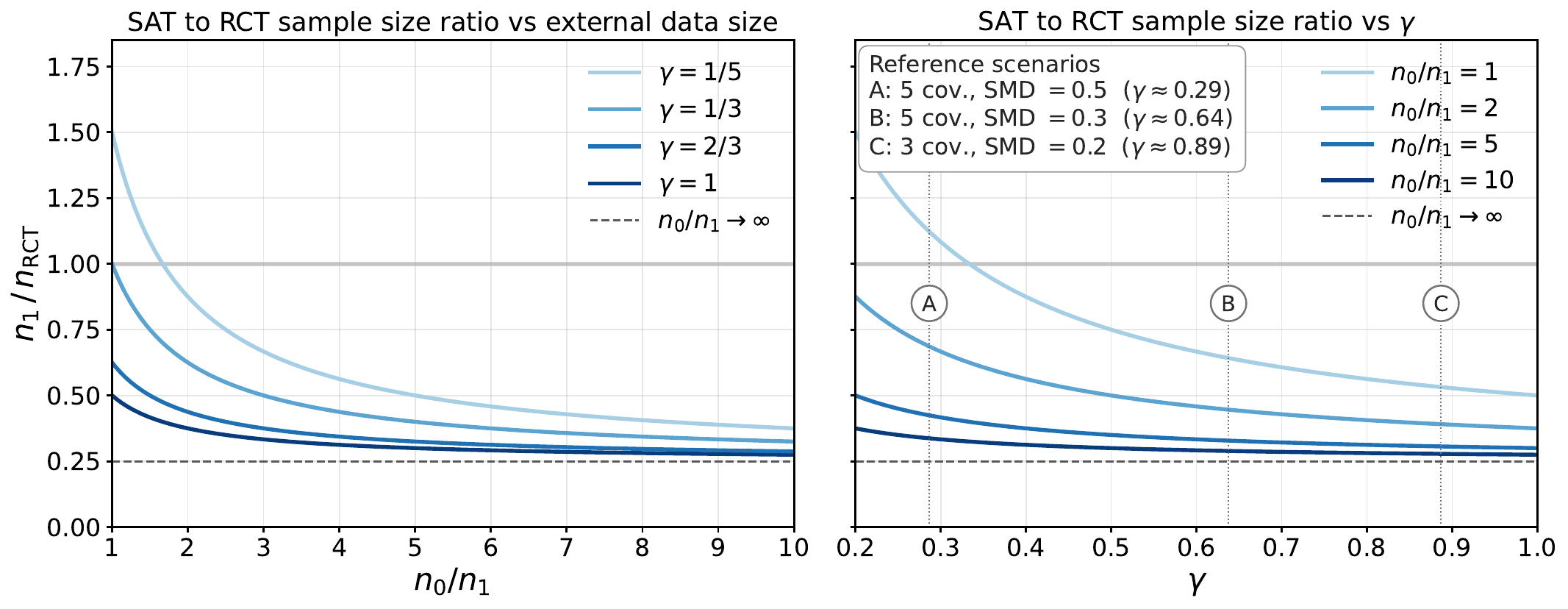}
    \caption{Ratio of the single-arm trial sample size (treated
    participants only) to the total sample size of a 1:1 randomized
    controlled trial with matched power. Left panel: ratio as a function
    of the external data size in units of the single-arm sample size, for
    several values of $\gamma$; the dashed line marks the $n_0/n_1 \to
    \infty$ limit, which would correspond to using the outcome-model plug-in estimator. Right panel: the same ratio as a function
    of $\gamma$ for several $n_0 / n_1$. Reference scenarios A, B, C
    correspond to $\gamma \approx 0.29, 0.64, 0.89$, computed from
    the Gaussian-SMD heuristic of Eq.~\eqref{eq:lambda-smd} under
    independence.}
    \label{fig:sample-size}
\end{figure}

\paragraph{Comparison to an RCT with matched power.}
It is useful to unpack what
$\gamma$ implies for sample size relative to a randomized comparator.
For a 1:1 RCT with total enrollment $n_{\mathrm{RCT}}$, the asymptotic
variance of the treatment effect estimator is $4\sigma^2 /
n_{\mathrm{RCT}}$, where $\sigma^2$ is the outcome marginal variance, assumed to be the same in the two arms under constant treatment effect. Matching power between the single-arm design and this
RCT amounts to equating variances, and with $\rho_0 = 0$ gives the sample size ratio
\begin{equation}
\frac{n_1}{n_{\mathrm{RCT}}} \;=\;
\frac{1}{4}\left(1 + \frac{n_1}{\gamma\,n_0}\right).
\label{eq:sample-size-ratio}
\end{equation}
Fig.~\ref{fig:sample-size}
illustrates Eq.~\eqref{eq:sample-size-ratio} from two complementary
perspectives. The left panel shows the ratio as a function of
$n_0 / n_1$, with curves for several values of $\gamma$. The right
panel shows the same ratio as a function of $\gamma$ itself, with
curves for several values of $n_0 / n_1$, together with three reference
scenarios of covariate imbalance.

The asymptotic limit of Eq.~\eqref{eq:sample-size-ratio} as
$n_0/n_1 \to \infty$ is $1/4$, which also coincides with the outcome-model plug-in limit:
when the outcome model is treated as an oracle, control-arm noise
vanishes entirely and the trial-arm sample size required for matched
power is exactly a quarter of that of a 1:1 RCT. This factor of $1/4$
can be understood as follows. The first $1/2$ comes
from not having to enroll a randomized control arm: for the same total
enrollment, the single-arm trial can devote all patients to the treated
arm rather than splitting them with controls. The second $1/2$ comes
from eliminating the control-arm sampling variance: in an RCT,
randomized controls contribute a $\sigma^2 / n_0$ term, whereas an
oracle outcome model (or an infinite external pool) provides noiseless conditional 
counterfactuals.

\paragraph{Monotonicity of effective sample size under data augmentation.}
We close this section noting a monotonicity property of
Eq.~\eqref{eq:aipw-variance-power}: adding any external control
data---however marginally relevant, as long as positivity continues to
hold---never decreases the effective control sample size
$n_{0}^{\mathrm{eff}} = \gamma n_0$. Let $P_1$ denote the trial covariate
distribution and $P_{0, A}$ the current historical pool's distribution, with
sample size $n_A$, and assume $P_{0,A}(x) > 0$ on the support of $P_1$. Consider the inverse efficiency factor
\begin{equation}
\gamma_A^{-1} \;=\; \int \frac{P_1(x)^2}{P_{0,A}(x)}\, dx.
\end{equation}
Now suppose we augment the pool by $n_B$ patients drawn from some
distribution $P_{0,B}$, so that the combined pool has size $n_0 = n_A + n_B$
and distribution
\begin{equation}
P_0(x) \;=\; \alpha\, P_{0,A}(x) + (1-\alpha)\, P_{0,B}(x),
\qquad \alpha \;=\; n_A/n_0.
\end{equation}
The combined inverse efficiency factor is $\gamma^{-1} = \int P_1^2 / P_0\, dx$.
Since $(1-\alpha)\, P_{0,B}(x) \ge 0$ everywhere, we have the pointwise
bound $P_0(x) \ge \alpha\, P_{0,A}(x)$, and therefore
$
\frac{P_1(x)^2}{P_0(x)} \;\le\;
\frac{1}{\alpha}\,\frac{P_1(x)^2}{P_{0,A}(x)}.
$ Integrating over $P_1$ support and substituting $\alpha = n_A/n_0$, we obtain:
\begin{equation}
\gamma n_0 \;\ge\; \gamma_A n_A.
\label{eq:nef-monotone}
\end{equation}
The inequality is strict whenever $P_{0, B}$ places positive mass on a
subset of $P_1$ support of positive measure, and reduces to
equality only when $P_{0, B}$ is supported entirely outside the trial's covariate space. Equivalently, every additional patient with nonzero propensity contributes to efficiency, while patients completely outside the trial population contribute nothing but also, in principle, do no harm.

Note that adding less relevant data typically \emph{decreases}
$\gamma$, because the average density ratio grows where $P_{0,B}$ dilutes
$P_0$ below $P_{0,A}$; however, the sample size $n_0$ grows proportionally faster, and so the effective sample size strictly increases. The
practical implication is that, from the standpoint of estimator
efficiency alone, there is no downside to broader historical
cohorts---provided positivity is maintained. The trade-offs that
motivate restricting the historical dataset in practice---ignorability,
cost, model misspecification, and data quality---are discussed in
Section~\ref{sec:historical-data}.

\section{External Data and the Outcome Model}
\label{sec:external-data}

Outcome-model-based approaches can use external data in two distinct
roles: the \emph{training data}, used to fit the outcome model, and the
\emph{historical control data}, used in the estimator itself through the
explicit sum over historical controls. Which roles are active depends
on the estimator. The AIPW estimator uses both: it requires a training
dataset to fit the outcome model, and a historical
dataset that enters the estimator directly through the augmentation
term. The outcome-model plug-in estimator uses only external data to train the outcome model, and the estimator itself depends on
the external population only implicitly, through the model predictions
evaluated on the trial covariates. In what follows we discuss both
datasets, beginning with the historical control data---which is subject to the
stronger constraints and is relevant specifically to AIPW (and other propensity-based estimators)---and then
turning to the training data.

\subsection{The Historical Control Dataset}
\label{sec:historical-data}

The ideal historical control dataset matches the trial population along the
dimensions relevant to the estimand. Concretely, it should:
\begin{enumerate}
    \item Measure all confounders required by the ignorability assumption, so that conditional exchangeability between the trial and external control populations can be defended.
    \item Have sufficient covariate overlap with the trial population to satisfy positivity.
    \item Record the outcomes of interest on a schedule consistent with the trial protocol, using comparable instruments and follow-up windows.
    \item Be collected in a manner that minimizes systematic biases, including secular trends, differential loss to follow-up, selection effects, and changes in measurement standards over time.
\end{enumerate}
Beyond these, the dataset must clearly capture outcomes under the
control condition defining the estimand. The size and relevance of
the dataset are then the remaining levers that practitioners must weigh.
A typical workflow for assembling the historical dataset might involve:
\begin{enumerate}
    \item Identifying and acquiring an initial batch of data relevant to the study.
    \item Analyzing that data and considering whether additional acquisition is warranted.
    \item Selecting the historical dataset as a subset of the available data.
\end{enumerate}
This process raises several natural questions: what factors should
govern the definition and acquisition of the initial dataset, and what
determines whether augmentation is worthwhile? How should the working
subset be chosen? What processes ensure trustworthy results?
The next two paragraphs discuss the trade-offs in historical data collection and operational considerations are discussed in \Sec{sec:data-handling}.

\paragraph{What factors encourage us to collect a larger dataset?}
The historical dataset must be broad enough that the support of the
trial covariate distribution is contained in the support of the
historical distribution, so that positivity is satisfied. Once this is
met, the only additional motivation for a larger dataset is to increase
estimator efficiency and statistical power. The contribution of any
additional data to power can be estimated using
\Eq{eq:aipw-variance-power}, provided its expected propensity for the
trial can be characterized. Note that only data with nonzero propensity
contributes. In practice,
there may be substantial amounts of data with low but nonzero
propensity, and as shown in \Sec{sec:power},
adding any such data \emph{always} strictly increases the effective
sample size. The gain may be marginal---particularly for data with very
low propensity---and may not justify the costs discussed below.

\paragraph{What factors encourage us to collect a smaller dataset?}
Broader datasets built from
heterogeneous sources, varying collection protocols, or different periods
of clinical practice are more prone to introducing unmeasured
confounders and measurement inconsistencies that violate ignorability.
Careful understanding of the data---including the clinical and
operational drivers of disease course in the indication---is an
essential guard against unmeasured confounding. Cost is also a
significant practical constraint. In the second step of the workflow above, the question is whether the efficiency gain from additional data
justifies its acquisition cost, which can be estimated quantitatively
using \Eq{eq:aipw-variance-power} and the expected propensity of the
marginal additions.

Given these competing considerations, a natural strategy is to collect a
dataset large enough to ensure positivity and adequate power, but no
broader, so as to limit the accumulated cost of data acquisition and the risk of violating ignorability by introducing unmeasured confounders.

\subsection{The Training Dataset}
\label{sec:training-data}

The data used to train the outcome model is subject to weaker formal
constraints than the historical data, because the estimator does not
depend on the training dataset directly---its influence is mediated
entirely by the outcome model. The constraints on the training data are
therefore inherited through the constraints the outcome model must
satisfy. In particular, the training data must still measure the
confounders that the outcome model is required to adjust for, so that
the model can represent the relevant conditional outcome distributions
on both the trial and the historical populations.

For the outcome model to be well-specified on both the trial and
historical datasets used in the estimator, it is highly preferable for
the training data to cover the same populations as those two datasets,
and broader training datasets are often beneficial. This is especially
true for complex outcome models: as discussed in
\Sec{sec:digital-twins}, learning from larger datasets and transfer learning can
materially improve generalization on the trial and historical
populations, reducing the risk of misspecification. This ability to
leverage broader training data is a key practical advantage of
outcome-model-based approaches over pure propensity-based alternatives, which
are more tightly coupled to the specific trial and external control
populations.

When overall data availability is limited, the historical data may need
to be a subset of (or equal to) the training data. In that case, the
outcome model must be fit with cross-validation or nested
cross-validation, so that residuals in the estimator come from
models not trained on the same participants used to estimate the residual.

\subsection{Data Handling}
\label{sec:data-handling}

The validity of any model-based analysis in a single-arm trial depends
as much on the handling of the data and the analytic workflow as on the
formal properties of the estimator. The following practices are
especially important in this setting.
\paragraph{Pre-specification.}
The statistical analysis plan should specify the estimator, outcome
model, propensity model, covariate set, missing-data strategy, and
inference procedures prior to observing the trial data. For machine-learning
models, pre-specification should typically cover the training data, model class,
hyperparameter search strategy, and procedures for any retraining or
recalibration involving trial outcomes. Pre-specification protects the analysis from post-hoc decisions that can introduce bias and is essential for regulated applications.
\paragraph{Data quality and provenance.}
Historical data should be audited for consistency, outlier
prevalence, missingness patterns, and provenance ensuring the conditions described in \Secs{sec:historical-data}{sec:training-data} are met and potential issues are identified early.
\paragraph{Separation of training and estimation data.}
When the training data and the historical data overlap, model
predictions on historical patients must be generated under
cross-validation or an equivalent held-out scheme, to avoid overfitting
bias in the estimator. Disjoint training and historical datasets are
preferable when feasible; when they are not, nested cross-validation
provides a principled alternative at the cost of additional computational overhead.
\paragraph{Data and model locking.}
When feasible, the outcome model used in the estimator should be fully
fixed before trial outcomes become available. This allows for validation of the model and a more robust assessment of credibility.  When a model is fit wholly or in part on the outcome data in the trial, by necessity the model is not final until after database lock and care must be taken to avoid data leakage.  Additionally, if concerns about the outcome model properties exist, they cannot be evaluated except in a \emph{post hoc} fashion or, when available, via testing on similar historical data.

Outcome models used in the plug-in and AIPW ATT estimators can be pre-specified and locked. Some estimators incorporate trial outcomes into the outcome model by design. In such cases the final outcome model cannot be
locked before trial readout, but the \emph{update procedure} should itself be fully pre-specified, so that no
investigator-driven decisions depend on the observed trial outcomes. The same considerations apply to the propensity model, with the difference that it only involves baseline covariates and can typically be built on that data before outcomes are completely recorded. Training data and code versions should be archived and made available to allow independent reproduction of the analysis.

\section{FDA Guidance and the Role of Outcome Models}
\label{sec:fda-guidance}

Most single-arm trials are not registrational; they are conducted at
the sponsor's discretion as earlier-phase or post-approval studies. For
those that are registrational, single-arm trials with digital twins
will interface with the recent FDA draft guidance on AI in drug
development, \emph{Considerations for the Use of Artificial Intelligence 
To Support Regulatory Decision-Making for Drug and Biological Products
}~\cite{FDA-AI-Regulatory-2025}, which outlines a risk-based framework
for the use of AI models in regulatory decision-making. We find this
guidance useful also for non-regulated applications that carry
nontrivial model risk and recommend that sponsors consider its
principles in such settings. A closely related FDA draft guidance on
externally controlled trials~\cite{fda2023externalcontrols} addresses the
design and conduct of trials that rely on external comparator data and
provides complementary considerations that apply regardless of whether
a digital twin is used.

The AI guidance is organized around the concept of \emph{context of
use}, a common principle in regulated applications. The central
question is how the AI model---in our setting the outcome model, and
potentially the propensity model when its complexity is
nontrivial---will be used, which in turn defines the risk profile. The
guidance proposes a seven-step framework for sponsors to evaluate model
credibility under the context of use. At a high level, the steps are:
\begin{enumerate}
    \item Define the question of interest. Clearly articulate the specific decision or question the AI model aims to address.
    \item Define the context of use. Outline the role and scope of the AI model within drug development, including how its outputs will influence study design or decision-making.
    \item Assess the AI model risk. Evaluate the risk along two dimensions: the consequence of wrong decisions about the question of interest, and the model's influence on those decisions.
    \item Develop a plan to establish AI model credibility within the context of use. Describe the model development and evaluation process in relation to the intended application.
    \item Execute the plan.
    \item Document the results and discuss any deviations.
    \item Determine the adequacy of the AI model for the context of use.
\end{enumerate}
In what follows we focus on steps 1--3 in the context of the AIPW
estimator and provide recommendations for assessing model credibility.
We apply the framework to the outcome model,
assuming for simplicity that the propensity model complexity is low; in
principle the same framework can be applied to the propensity model when it 
is sufficiently complex.

In a single-arm trial analyzed with AIPW, the question of interest is
the standard efficacy question: whether the treatment effect over a
defined standard of care is statistically significant for one or more
endpoints. In early-phase trials, efficacy may be secondary to safety,
pharmacokinetic, and pharmacodynamic objectives, but it remains the
central question that the digital twin synthetic controls are used to address.

The context of use of the digital twin model is to predict outcomes under a
defined standard of care for the endpoints of interest, which then enter the AIPW estimator. 
Importantly, that means that the outcome model will be used in conjunction with 
a propensity model and historical data to estimate treatment effects.  This is a different 
context of use than the outcome-model plug-in estimator, in which model predictions are directly compared 
to observed outcomes to estimate treatment effects.

Choosing AIPW is itself a commitment to obtaining a robust and reliable treatment
effect estimate, which in turn requires that the ingredients of the
estimator are in place: a historical dataset that plausibly satisfies
positivity and ignorability, and the ability to construct at least one
of the nuisance models---outcome or propensity---to a sufficient
standard. These elements are central to the risk assessment.

The FDA guidance decomposes model risk into two dimensions:
\emph{decision consequence} and \emph{model influence}. Decision
consequence is a model-independent statement about the impact of a
wrong decision on the question of interest; model influence is the
extent to which the model determines that decision within the context
of use.

Decision consequence is invariably high in the registrational
single-arm setting: bias in the treatment effect estimate or its
variance translates directly into an incorrect efficacy conclusion. 
This can impact both incorrect types of decisions: wrong decisions can mean 
that ineffective treatments can appear effective, and effective treatments 
can appear ineffective.  For non-registrational, early phase studies we might regard the 
decision consequence to be lower as the factors influencing the continuation or 
stopping of the program can depend on other factors such as safety, 
pharmacokinetics, and pharmacodynamics.

This motivates careful assessment of the estimator. As discussed above,
we should have justification that the historical data satisfies the
causal inference assumptions required for external
comparison~\cite{fda2023externalcontrols} and that at least one nuisance
model can be well-specified. This confidence comes primarily from the
data itself. When historical data come from highly relevant
sources---recent clinical trials or observational studies in the same
population---or when real-world data can be validated to yield outcomes
consistent with standard-of-care expectations in the trial population,
there is a defensible basis for the estimator and for model
development.  We can regard this as an evaluation of the credibility 
of the approach that exists outside of the model itself.

\begin{figure}
    \centering
    \includegraphics[width=0.5\linewidth]{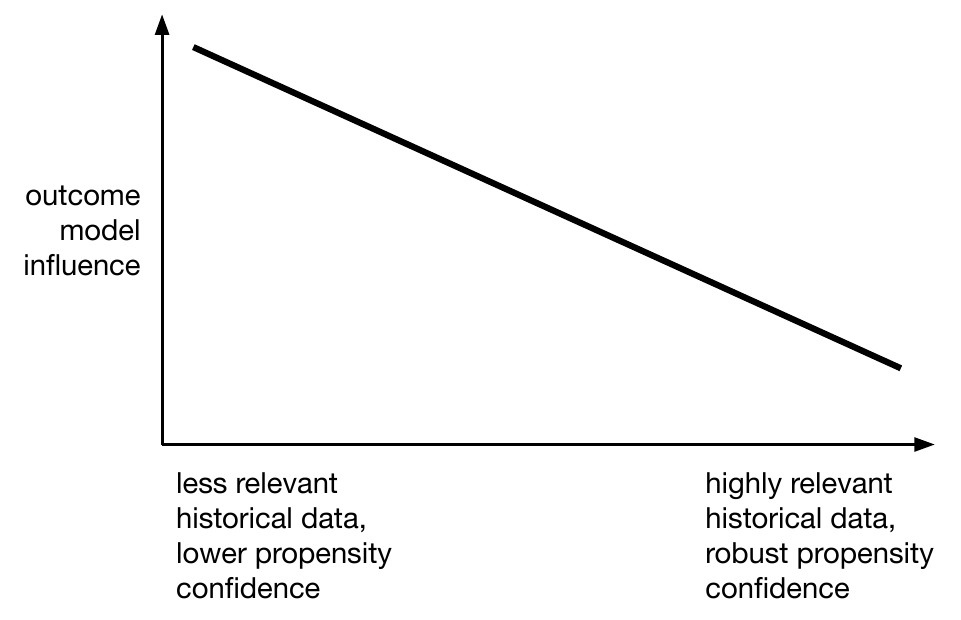}
    \caption{Outcome model influence is lowest when highly relevant historical data and a well-characterized propensity model allow the causal inference assumptions to be confidently satisfied. As relevance or propensity knowledge declines, the analysis relies increasingly on the outcome model being correctly specified, and the model's influence on the treatment effect estimate grows.}
    \label{fig:aipw_model_influence}
\end{figure}

Model influence, by contrast, depends sensitively on the relevance of
the historical data and on the quality of the propensity model. When
the causal inference assumptions are confidently satisfied and the
propensity model is well-specified, the outcome model's influence on the
treatment effect estimate is low, because AIPW remains consistent under
correct propensity specification alone. This regime is typical when
patient-level data are available from recent trials in the same
indication, with identifiable factors governing study inclusion (such
as severity or biomarker criteria) and a clear effect of those factors
on disease progression. In such cases, IPW on matched historical data
may itself suffice to obtain a consistent estimate, with AIPW adding efficiency, robustness, and enabling easier inference, if the outcome model is also well-specified. When these conditions do not hold---the historical data is less relevant (older, or from trials with shifted
inclusion criteria), or the propensity is poorly characterized---the
reliance on the outcome model to be correctly specified increases, and
so does its influence. This trade-off is illustrated in
\Fig{fig:aipw_model_influence}.

In the highest-influence regime, the outcome model carries essentially
the entire burden of identification. That would correspond to the case where relevant, unconfounded historical patient-level data is scarce or not directly available. When this is the case, one might consider an outcome-model plug-in estimator as the AIPW bias correction cannot be computed reliably. Of course, if unconfounded historical patient-level data is scarce or missing, this increases scrutiny on the data used for model training and most importantly on validation of model predictions and whether any assumptions integrated in the model are supported by evidence and are defensible. This would be the case for the examples mentioned in \Secs{sec:strategy}{sec:digital-twins} where models are used to overcome lack of direct identifiability in historical patient-level data.

We note that the model influence may change depending on whether 
other synthetic control approaches are used in the study design.  We may also compute treatment 
effects using propensity methods our the outcome-model the plug-in estimator.  In cases where 
a propensity method is used as the primary endpoint, the model influence is lowered 
substantially.  In cases where the plug-in is used as the primary endpoint, it 
is raised.

\subsection{Considerations from the Real World Evidence Guidance}
\label{sec:rwe-guidance}

As already mentioned, additional considerations are provided in the FDA guidance on externally controlled trials~\cite{fda2023externalcontrols}, especially for data-based control comparators. This guidance emphasizes several concepts discussed here, including prespecification of key components such as data sources, selection criteria for comparator data, endpoint definitions, modeling approaches, and the overall analysis plan. Technical challenges, such as data traceability and the definition of time zero, are also important, as are steps to avoid data leakage in propensity and outcome models. In addition, sensitivity analyses are recommended to evaluate the robustness of the external control. Additional guidance documents may apply depending on the type of data used, such as electronic health records~\cite{fda2024ehrclaims} or registries~\cite{fda2023registries}, which provide domain-specific considerations for the use of these data.

\section{Case Studies}
\label{sec:case-studies}

We illustrate the methods discussed in the previous sections through
reanalyses of two randomized controlled trials: the Celebrex trial in
Amyotrophic Lateral Sclerosis (ALS) \cite{cudkowicz-2006} and the
2CARE trial in Huntington's disease (HD) \cite{mcgarry-2017}. In each
reanalysis, we treat the trial arms as though they were single-arm
studies, construct external comparators from historical control data and/or digital twin models,
and estimate the treatment effect using the four estimators reviewed
in \Sec{sec:estimators}: propensity score matching (PSM), inverse probability weighting (IPW), the outcome-model plug-in estimator
(OM), and augmented inverse probability
weighting (AIPW). For each trial, we perform
two analyses:
\begin{enumerate}
    \item \textbf{RCT-control vs.\ external control.} The original randomized control arm is treated as a single-arm trial and compared against an external control arm. Under valid causal inference, the estimated treatment effect should be zero.
    \item \textbf{RCT-treated vs.\ external control.} The original treated arm is treated as a single-arm trial. The true treatment effect is unknown, so we report bias relative to the RCT-internal treatment effect estimate; the variance of the difference accounts for the variance of the RCT estimate as well.
\end{enumerate}
In both analyses we consider multiple clinical endpoints
and report the \emph{absolute standardized bias}---the absolute difference between estimated and expected treatment effect, divided by the outcome standard deviation in
the external control population with matched eligibility criteria---together with 95\% confidence
intervals. We also report the fractional reduction of the treatment
effect estimator's variance relative to one-to-one PSM, as a measure of the
efficiency gained by alternative methods. 

Details about the trial, historical, and training datasets are provided in Appendix~\ref{app:external-data}, along with the standard deviations used to compute standardized bias. A description of the propensity and 
digital twin models used for this analysis are provided in Appendix~\ref{app:digital-twin-models}. All analyses share the following methodology. PSM uses nearest-neighbor one-to-one matching with replacement with a
caliper computed as $0.2\times\mathrm{SD}(\mathrm{logit}(\hat{e}))$~\cite{austin2011caliper}. For
IPW and AIPW, we restrict the analysis to the overlap region of the
propensity distribution; this stabilizes the propensity odds and
ensures that the estimator is evaluated where positivity is
well-supported.

Variances and confidence intervals are constructed via bootstrap. For
PSM, IPW, and OM the nuisance models are refit on
each bootstrap sample. For AIPW the nuisance models are held fixed,
as AIPW is insensitive to leading-order nuisance estimation error
by construction. For PSM we use the
subsampling bootstrap of Politis and Romano
\cite{politis-1994}, which is consistent for the matching
estimator \cite{abadie-2008}. From the bootstrap variance
$\hat V$, we construct 95\% confidence intervals using a normal
approximation $\hat\tau \pm 1.96\sqrt{\hat V}$. 

\subsection{Amyotrophic Lateral Sclerosis: Celebrex}
\label{sec:celebrex}

The Celebrex trial \cite{cudkowicz-2006} was a phase II, randomized,
double-blind, placebo-controlled trial of the COX-2 inhibitor
celecoxib in approximately 300 patients with ALS. The primary
objective was to evaluate whether celecoxib slows disease progression,
as measured by the rate of decline in the revised ALS Functional
Rating Scale (ALSFRS-R) over 12 months. The trial did not demonstrate
a benefit of celecoxib on the primary or secondary endpoints.

\begin{figure}[t!]
    \centering
    \includegraphics[width=0.90\linewidth]{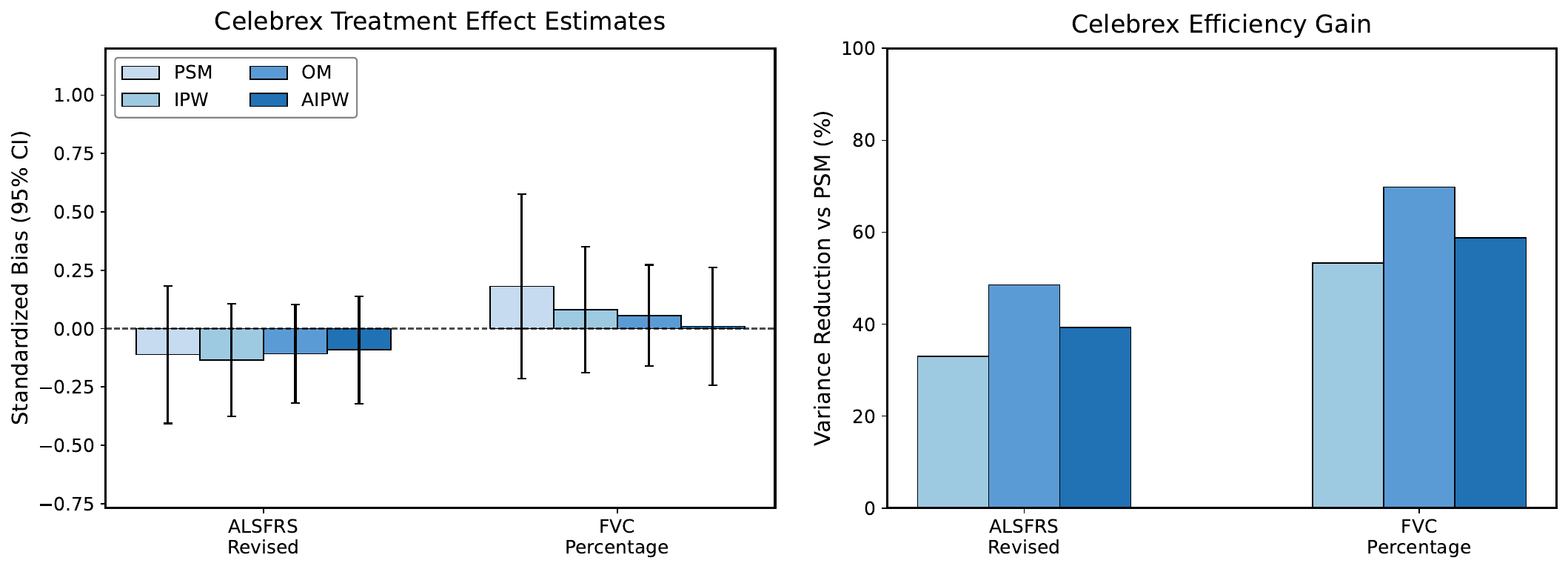}
\includegraphics[width=0.90\linewidth]{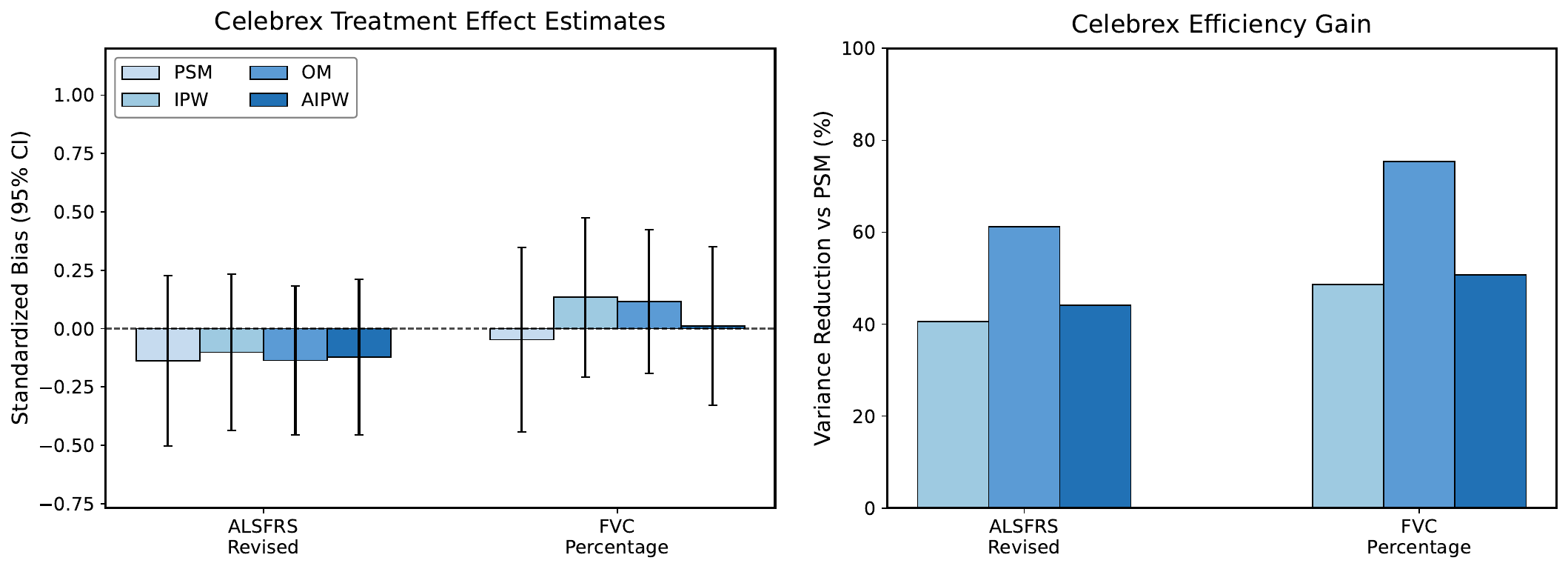}
    \caption{Celebrex analysis. Top: control arm as a single-arm study vs.\ external control. Left: absolute standardized bias per endpoint, across estimators, with 95\% bootstrap confidence intervals. We compare one-to-one propensity score matching (PSM), inverse probability weighting (IPW), outcome-model plug-in (OM), and augmented inverse propbability weighting (AIPW). Right: fractional variance reduction of each estimator relative to one-to-one PSM. Bottom: treated arm as a single-arm study vs.\ external control.}
    \label{fig:celebrex}
\end{figure}

For this reanalysis we consider two endpoints at 12 months, the
primary analysis timepoint of the trial: ALSFRS-R total score and
forced vital capacity as a percentage of predicted (FVC\%). Results for
the control and treated arms are shown in \Fig{fig:celebrex}. Each figure reports, in the left
panel, the absolute standardized bias for each estimator and endpoint
with 95\% confidence intervals; the right panel reports the
fractional variance reduction relative to one-to-one PSM.

Averaged across arms and endpoints, AIPW exhibits the smallest absolute
standardized bias, approximately $6\%$, compared to $11\%$ for
the other estimators. These differences should be interpreted with
caution: they represent point estimates from a single sample, and the
confidence intervals cover zero for all methods on all endpoints.
Variance reduction relative to one-to-one PSM is substantial across the board.
IPW alone yields a significant reduction relative to PSM, driven by
its use of the full historical dataset in the overlap region rather
than retaining a single matched control per treated participant. The
OM and AIPW estimators produce an additional reduction, reflecting the outcome model's ability to explain part of the outcome variance.

\subsection{Huntington's Disease: 2CARE}
\label{sec:2care}

The 2CARE trial \cite{mcgarry-2017} was a phase III, randomized,
double-blind, placebo-controlled trial of coenzyme Q10 (CoQ10) in approximately 600 patients with early-stage Huntington's
disease. The primary objective was to evaluate whether CoQ10 slows
disease progression, measured by change in the Total Functional
Capacity (TFC) subscale of the Unified Huntington's Disease Rating
Scale (UHDRS) over 5 years. The study was discontinued early for
futility following a planned interim analysis. For the reanalysis we
therefore evaluate endpoints at 4 years, where there is still sufficient
data availabile.

\begin{figure}[t!]
    \centering
    \includegraphics[width=0.90\linewidth]{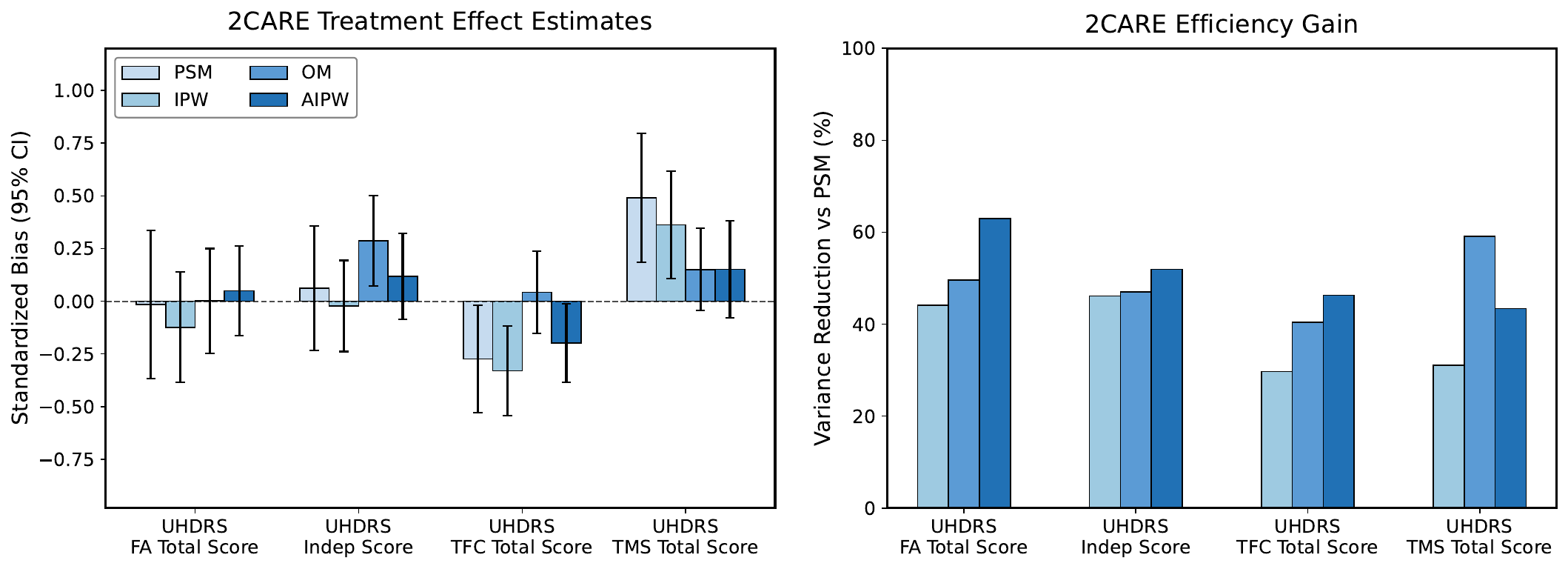}
    \includegraphics[width=0.90\linewidth]{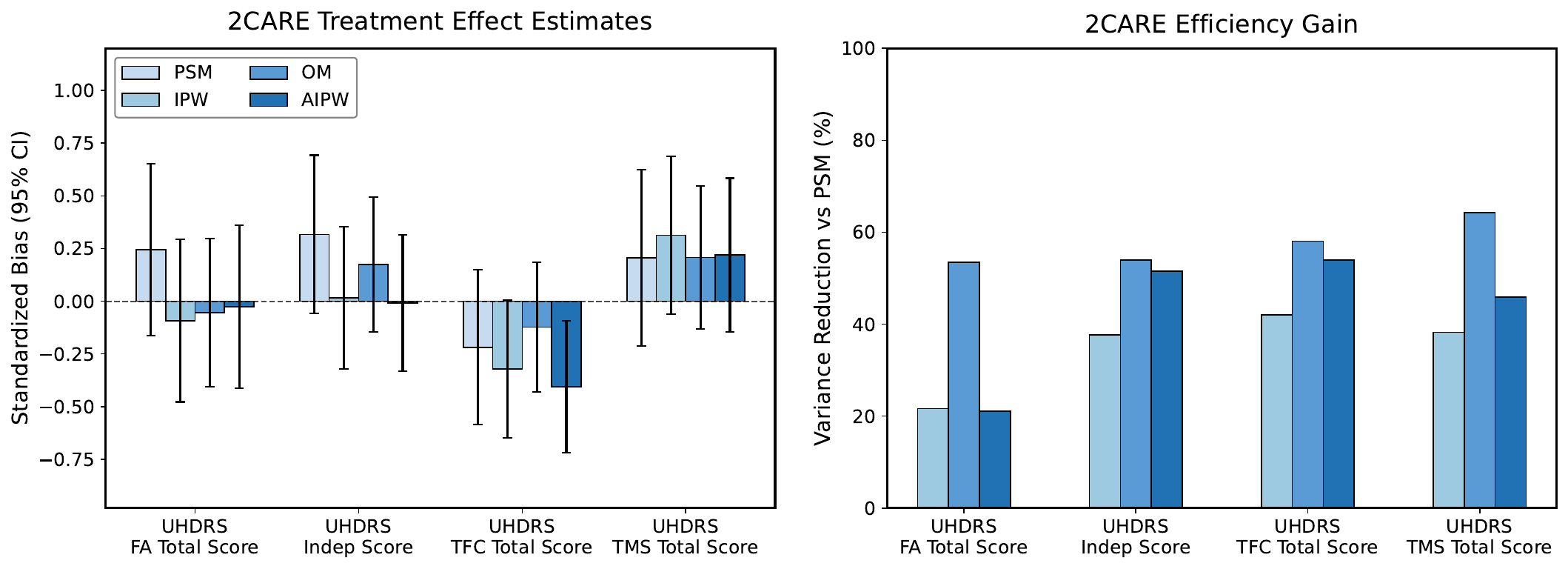}
    \caption{2CARE analysis. Top: control arm as a single-arm study vs.\ external control. Panels and conventions as in \Fig{fig:celebrex}. Bottom: treated arm as a single-arm study vs.\ external control.}
    \label{fig:2care}
\end{figure}

We consider four endpoints from the UHDRS: the Functional Assessment
Total Score, the Independence Score, the Total Functional Capacity
Total Score, and the Total Motor Score. Results for the control and
treated arms are shown in \Fig{fig:2care}.

Averaged across arms and endpoints, AIPW's absolute standardized bias is comparable to that of the outcome-model plug-in estimator (approximately $14\%$), and smaller than the other methods (approximately in the range $20$ - $23\%$). As in the ALS reanalysis, most
confidence intervals cover zero on.
Variance reduction relative to one-to-one PSM is again substantial, with the
largest gains concentrated in the model-based estimators, consistent
with the pattern observed in the Celebrex reanalysis.

\subsection{Interpretation}
\label{sec:case-studies-discussion}

Two consistent patterns emerge across the two case studies. First, all four estimators produce point estimates of the standardized
bias that are small and in the vast majority of cases not significantly different from zero. In both the
ALS and HD reanalysis, AIPW's absolute bias point estimate is on average smaller than those
of the other methods. These findings are consistent with the double-robustness property of AIPW, which provides additional protection against misspecification of either the outcome or the propensity model, but they do not by themselves constitute conclusive
evidence of superior bias control given the confidence intervals and
the single-sample nature of the comparison.
Second, all methods consistently increase efficiency compared to one-to-one PSM. Moving from one-to-one PSM to IPW produces a meaningful
reduction in variance, driven by IPW's use of the entire historical
dataset in the overlap region. Moving further to plug-in and AIPW
produces an additional reduction, reflecting the increased efficiency and the variance absorbed by
the outcome model via the $(1-\rho_0^2)$ factor of the conditional variance in
\Eq{eq:aipw-variance-power}. 

Taken together, these case studies support the conclusion that model-based approaches, and AIPW in particular, offer
favorable estimators in realistic settings: they match or
improve on the bias control of pure propensity-based methods while
delivering substantial and consistent efficiency gains. 

\subsection{Simulating Reduced Overlap}
\begin{figure}[t!]
    \centering
    \includegraphics[width=0.95\linewidth]{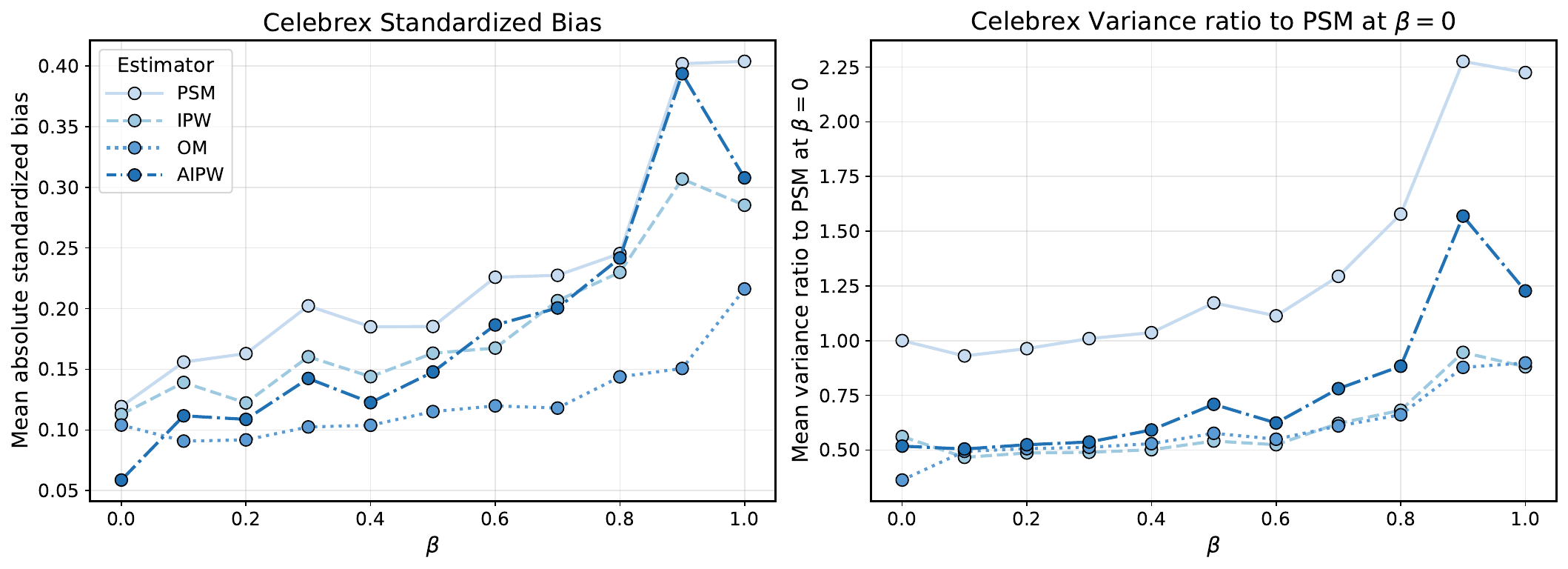}
    \caption{Overlap-resampling sweep for the ALS Celebrex analysis.
    Left: mean absolute standardized bias as a function of $\beta$,
    averaged over endpoints, arms, and resampled datasets. Right: mean
    estimator variance as a function of $\beta$, normalized by the
    PSM variance at $\beta = 0$, and averaged over endpoints, arms, and resampled datasets.
    $\beta = 0$ corresponds to the unmodified historical pool used
    in the main analysis of \Sec{sec:celebrex}; larger $\beta$ corresponds to reduced
    overlap with the trial covariate distribution.}
    \label{fig:als-resampling}
\end{figure}

\begin{figure}[t!]
    \centering
    \includegraphics[width=0.95\linewidth]{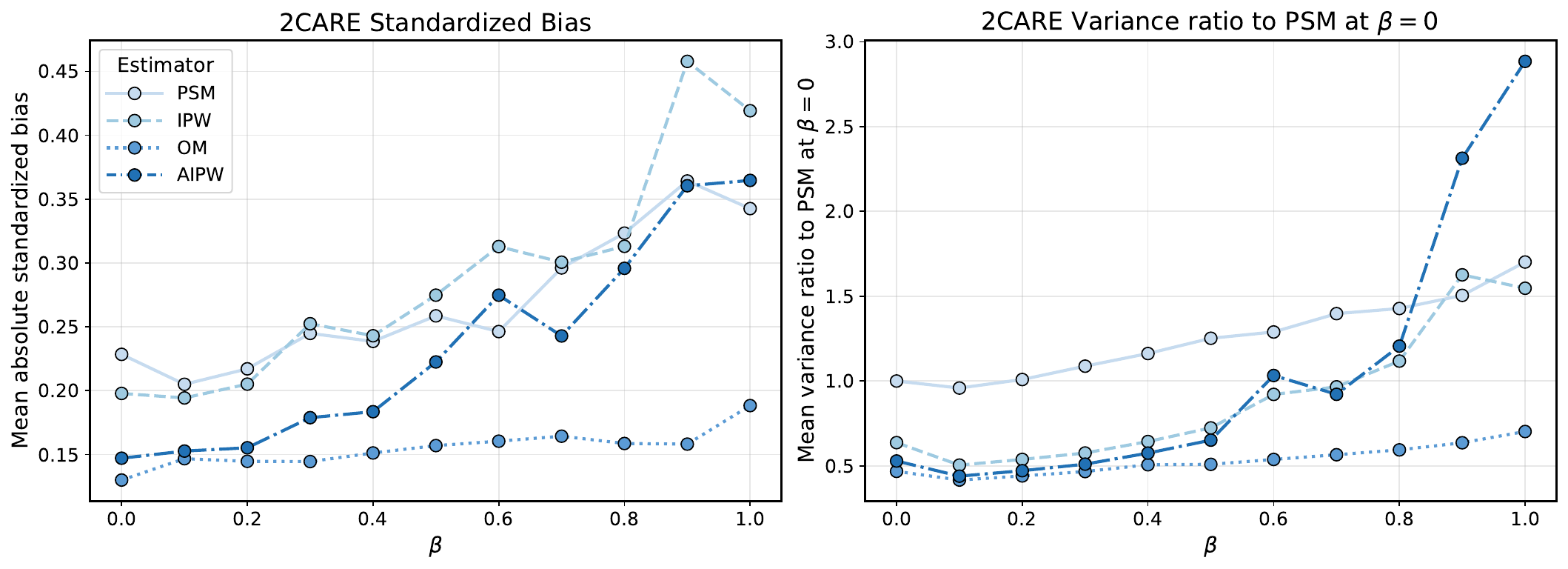}
    \caption{Overlap-resampling sweep for the HD 2CARE analysis.
    Panels and conventions as in Fig.~\ref{fig:als-resampling}.}
    \label{fig:hd-resampling}
\end{figure}

To probe how each estimator behaves as the effective overlap between
the trial and historical control populations degrades, we perform an
\emph{overlap-resampling} analysis. Starting from the full external 
dataset, we resample each subject with
probability proportional to
$\left[(1 - e(X))/e(X)\right]^{\beta}$, where $e(X)$ is the propensity
score for trial membership and $\beta$ is a tunable exponent. A
positive $\beta$ downweights subjects most similar to the trial,
thereby narrowing the overlap of the resampled pool with the trial
covariate distribution; $\beta = 0$ is equivalent to the
unmodified historical pool. The resampled pool
serves as the training set for the outcome model, and its subset satisfying eligibility criteria as the the historical dataset for propensity matching, weighting, or bias correction. For
each $\beta$, we draw $20$ independent resamples and re-run the full AIPW / IPW / OM / PSM pipeline on each. For simplicity, we estimate the variance using asymptotic formulas with fixed nuisance models. In Fig.~\ref{fig:als-resampling} we report results for the Celebrex study and in Fig.~\ref{fig:hd-resampling} for the 2CARE study. The bias panel reports the absolute standardized
bias as a function of $\beta$. For each value of $\beta$ and method, we average the absolute standardized bias over resamples of the external data and then over outcomes and arms. The variance panel reports the ratio of the variance of each estimator to the variance of the PSM estimator at $\beta=0$. For each value of $\beta$ and method, we compute the average of the ratio across resamples of the external data and then over outcomes and arms. The efficiency factor $\gamma$ also decreases with $\beta$ by construction.

Across both case studies,
the outcome-model-based estimators (OM and AIPW) maintain smaller or comparable 
mean absolute standardized bias than PSM and IPW at low or moderate $\beta$. As $\beta$ grows and the effective overlap
shrinks, the bias of all four estimators drifts upward, but OM
remains consistently the lowest. This is
consistent with the outcome model being able to draw on a broader
training pool that retains useful generalization to the trial
covariate region even when the within-trial-overlap subset of
historical data shrinks. The variance panels show the pattern
expected from the asymptotic analysis: at
$\beta = 0$ the model-based estimators (OM, AIPW) and IPW are
typically more efficient than PSM, and as $\beta$ grows the weighting-based estimators become progressively more unstable.
The OM estimator is the least sensitive, since its variance depends on the
outcome-model residual variance rather than directly on the weights.
Taken together, the two case studies indicate that
outcome-model-based estimators, and outcome-model plug-in in particular, can be
more robust to deterioration of historical overlap than purely
propensity-based estimators.


\section{Discussion}
\label{sec:discussion}

Single-arm trials are an important study design in modern drug
development. In
this work we have argued that outcome-model-based synthetic controls---in
particular those built with modern machine learning models---offer a
rigorous and practical complement to direct data-based methods, and
deserve a more central role than they currently occupy in the analysis
of single-arm trials.

We reviewed common ATT estimators for single-arm trials and articulated the expected benefits of outcome-model-based approaches. We derived power and sample size formulas for the AIPW estimator under homoskedasticity and a constant treatment effect and discussed practical ways to estimate sample size. On the operational side, we
articulated a framework for selecting external data, discussed the
distinct roles of training and historical datasets, outlined data-handling practices, and articulated recommendations to map the methodology onto the recent FDA draft
guidance on AI in drug development. Two reanalyses of completed randomized trials, the Celebrex
trial in ALS and the 2CARE trial in Huntington's disease, illustrate
the methods on realistic endpoints and sample sizes.

A few practical takeaways emerge. First, when patient-level historical control data that likely allows identification of the treatment effect, is available, outcome-model-based approaches should
be viewed not as a replacement for propensity-based methods but as a
layered extension of them: AIPW combines an outcome model with a
propensity model and is formally no less robust than either component
on its own, while achieving the semiparametric efficiency bound when
both are well-specified. In the case studies this manifests as
comparable or smaller point-estimate bias together with substantial
and consistent variance reduction relative to one-to-one PSM, typically in
the range of 40--60\% depending on endpoint and arm, with the
model-based estimators (plug-in and AIPW) achieving the largest gains.
Second, the theoretical analysis makes explicit what drives the
efficiency gain: moving from PSM to AIPW 
allows more efficient use of available data and captures the
outcome model's explanatory power. Because this second gain is contingent on a
well-specified outcome model, the case for AIPW is strongest when the
training data are rich and diverse enough to support flexible,
well-validated digital twins---a condition that is increasingly met as
historical datasets and modeling methods mature. 

Several natural extensions merit attention. Our discussion has focused on continuous
outcomes; time-to-event and binary endpoints follow the same
semiparametric logic but require adaptation of the efficient influence
function and care in the variance decomposition. The
non-identifiable setting---in which patient-level external data do not
directly represent the target control population but summary
statistics from historical trials are available---was raised but not
developed here, and is a setting in which model-based approaches offer particular
promise. 

More broadly, the emergence of digital twin models and the concurrent
codification of regulatory expectations, through the FDA's draft
guidances on AI in drug development and on externally controlled
trials, create a timely opportunity to modernize the statistical
practice of single-arm trial analysis. The methods and discussion in this paper are
intended to support that transition: to equip practitioners with tools
to design single-arm trials with informed power calculations, to
select external data in a principled way, and to deploy
digital twin outcome models within an estimator framework that carries
interpretable statistical guarantees and aligns with the credibility
framework that regulators are beginning to articulate. We view
outcome-model-based synthetic controls as a core component of the toolkit for ethical, efficient,
and evidence-generating single-arm trials in the coming years.

\bibliographystyle{unsrt}
\bibliography{bibliography}

\appendix
\section{IPW Estimator for ATT}
\label{app:ipw}
\paragraph{Importance sampling.}
The second term of the IPW estimator in \Eq{eq:ipw} can be derived
from an importance-sampling identification of the counterfactual mean
$\tau_0 \equiv \E[Y(0)\mid A=1]$. By Bayes' rule, the density ratio
between the treated and control covariate distributions is
\begin{equation}
\label{eq:density-ratio}
    r(X) \;\equiv\; \frac{P(X\mid A=1)}{P(X\mid A=0)}
    \;=\; \frac{1-p_1}{p_1}\,\frac{e(X)}{1-e(X)},
\end{equation}
where $p_1 = P(A=1)$ and $e(X) = P(A=1\mid X)$. Under ignorability,
$\E[Y(0)\mid X, A=1] = \E[Y\mid X, A=0] = \mu_0(X)$, so $\tau_0$ can
be written as an expectation over the control covariate distribution
reweighted by $r(X)$:
\begin{align}
\tau_0
    &= \E_{X\sim P(X\mid A=1)}\!\left[\mu_0(X)\right] \\
    &= \E_{X\sim P(X\mid A=0)}\!\left[r(X)\,\mu_0(X)\right] \\
    &= \E\!\left[r(X)\,Y \,\big|\, A=0\right].
\end{align}
The corresponding finite-sample estimator uses
$\hat p_1 = n_1/n$ and $1 - \hat p_1 = n_0/n$, so
$(1-\hat p_1)/\hat p_1 = n_0/n_1$. Substituting in $r(X)$, the
empirical version of $\tau_0$ simplifies to
\begin{equation}
    \hat\tau_0
    \;=\; \frac{1-\hat p_1}{\hat p_1}\cdot\frac{1}{n_0}
    \sum_{i:A_i=0}\frac{e(X_i)}{1-e(X_i)}\,Y_i
    \;=\; \frac{1}{n_1}\sum_{i:A_i=0}\frac{e(X_i)}{1-e(X_i)}\,Y_i,
\end{equation}
which is the second term of \Eq{eq:ipw}. Equivalently, the propensity
odds $w(X) = e(X)/(1-e(X))$ act as importance weights that map
expectations under the control covariate distribution
$P(X\mid A=0)$ to expectations under the treated covariate
distribution $P(X\mid A=1)$, with the population factor
$(1-p_1)/p_1$ absorbed by the $1/n_1$ normalization.

\paragraph{Asymptotic Variance.}
In this section, we show that the IPW estimator of $\tau_{ATT}$
defined in \Eq{eq:ipw} is asymptotically normal and derive its
asymptotic variance. Throughout, we
assume that the propensity odds $w(X) = e(X)/(1-e(X))$ are known or estimated
with error $o_p(n^{-1/2})$, so that for the purposes of the asymptotic
variance we can treat $w(X)$ as a known function of $X$. As noted in the main text, if the propensity model is parametric, propensity error can be included via M-estimation; alternatively bootstrap can be used. The goal of this section is to derive an  asymptotic formula under simple conditions, in order to compare the expected behavior against other estimators.

The IPW estimator in \Eq{eq:ipw} is the difference of two sample
averages computed on disjoint groups of i.i.d.\ observations. In the large-sample, limit each
is asymptotically normal by the central limit theorem, and so is
their difference,
\begin{equation}
\label{eq:ipw-asympt-normal}
\sqrt{n}\!\left(\hat{\tau}_{ATT}^{\text{IPW}} - \tau_{ATT}\right)
\xrightarrow{\;d\;}
\mathcal{N}\!\left(0,\;\sigma^2_{\text{IPW}}\right).
\end{equation}
The asymptotic variance equals to the sum of the contributions from
the treated and control groups,
\begin{equation}
\label{eq:ipw-sigma2}
\sigma^2_{\text{IPW}}
\;=\; \frac{\sigma_1^2}{p_1}
\;+\; \frac{1-p_1}{p_1^2}\,\V[w(X)Y\mid A=0],
\end{equation}
where $\sigma_1^2=\V[Y|A=1]$ and $n_1 = np_1$, $n_0 = n(1-p_1)$ in the large-sample limit. The variance of the estimator at sample size $n$ is given by:
\begin{equation}
\label{eq:ipw-var-from-sigma2}
\V\!\left[\hat\tau_{ATT}^{\text{IPW}}\right]
\;\approx\; \frac{\sigma^2_{\text{IPW}}}{n}.
\end{equation}
We have assumed that the contribution of weight-estimation error is $o_p(n^{-1/2})$ and therefore negligible relative to the leading
$O_p(n^{-1/2})$ sampling fluctuations. If we condition on $X$ inside the second expectation in
\Eq{eq:ipw-sigma2} and apply the law of total variance we obtain
\begin{equation}
\label{eq:ipw-inner-var}
\V[w(X)Y \mid A=0]
\;=\; \E\!\left[w^2(X)\,\kappa_0^2(X)\,\big|\,A=0\right]
\;+\; \V\!\left[w(X)\,\mu_0(X)\,\big|\,A=0\right],
\end{equation}
where $\kappa_0^2(X) = \V[Y\mid X, A=0]$, and we used that $w(X)$ is
constant under the conditional law of $Y\mid X, A=0$ and can be
pulled out of the inner conditional variance. Under homoskedasticity
the conditional outcome variance is constant in $X$,
$\kappa_0^2(X) = \kappa_0^2$, and the first term in
\Eq{eq:ipw-inner-var} reduces to $\kappa_0^2\,\E[w^2(X)\mid A=0]$. We can further assume that the conditional variance is equal in both treated and control groups and define $\kappa_0^2=\kappa_1^2=\kappa^2$. Recalling the definition of $n_0^{\mathrm{eff}}$ in \Eqs{eq:neff}{eq:gamma}, we obtain
\begin{equation}
\V\!\left[\hat\tau_{ATT}^{\text{IPW}}\right]
\;\approx\; \frac{\sigma_1^2}{n_1}+\frac{\kappa^2+\Delta}{n_0^{\mathrm{eff}}},
\end{equation}
where 
\begin{equation}
\Delta \;=\; \frac{\V[w(X)\mu_0(X)\mid A=0]}{\E[w^2(X)\mid A=0]}.
\end{equation}

\section{Efficient Influence Function for ATT}
\label{app:eif}
In this section we derive the efficient influence function for the average treatment effect on the treated, under standard identifiability assumptions. The form of the AIPW estimator for ATT can then be derived from the efficient influence function. The estimand can be written as:
\begin{equation}
\begin{aligned}
    \tau_{ATT} &= \E[\mu_1(X) - \mu_0(X) | A=1],
\end{aligned}
\end{equation}
where $\mu_a(X) = \E[Y|X, A=a]$. For simplicity, we define
\begin{equation}
\begin{aligned}
    \tau_1 &= \E[\mu_1(X) | A=1]\text{ and } \tau_0 = \E[\mu_0(X) | A=1],\\
\end{aligned}
\end{equation}
so that
\begin{equation}
\begin{aligned}
    \tau_{ATT} &= \tau_1 - \tau_0.
\end{aligned}
\end{equation}
\textbf{Step 1: rules and building blocks}. Instead of explicitly computing Gateaux derivatives, we will use the \emph{building blocks} strategy outlined in \cite{kennedy2023review} to compute the efficient influence function $\EIF[\cdot]$. In particular, we will use the following rules:
\begin{enumerate}[label=\alph*.]
    \item Pretend the data are discrete.
    \item Treat influence functions as derivatives and use standard differentiation rules:
    \begin{itemize}
        \item $\EIF[\psi_1 + \psi_2]=\EIF[\psi_1] + \EIF[\psi_2]$
        \item $\EIF[\psi_1\psi_2]=\EIF[\psi_1]\psi_2+\psi_1\EIF[\psi_2]$
        \item $\EIF[\frac{\psi_1}{\psi_2}]=\frac{\EIF[\psi_1]\psi_2-\psi_1\EIF[\psi_2]}{\psi_2^2}$
    \end{itemize}
    \item Use building blocks \cite{kennedy2023review, hines2022demystifying}: 
    \begin{itemize}
        \item $\EIF[\E[Z]] = Z - \E[Z]$,
        \item $\EIF\big[\mu_a(X)\big] = \frac{\mathbb{1}(X=x,A=a)}{p(x, a)}\left[Y - \mu_a(X)\right]$.
    \end{itemize}
\end{enumerate}
\textbf{Step 2: use Step 1 rules to rewrite $\EIF[\tau_{ATT}]$}. We have $\EIF[\tau_{ATT}] = \EIF[\tau_1] - \EIF[\tau_0]$, where 
\begin{equation}
\begin{aligned}
    \EIF[\tau_1] & = \EIF\left[\sum_x\mu_1(x)p(x|A=1)\right]\\
    & = \sum_x\left[\EIF[\mu_1(x)]p(x|A=1) +\mu_1(x)\EIF[p(x|A=1)]\right],\\
    \EIF[\tau_0] & = \EIF\left[\sum_x\mu_0(x)p(x|A=1)\right]\\
    & = \sum_x\left[\EIF[\mu_0(x)]p(x|A=1) +\mu_0(x)\EIF[p(x|A=1)]\right].\\
\end{aligned}
\end{equation}
\textbf{Step 3: use Step 1 rules to compute conditional density building block}.
\begin{equation}
\begin{aligned}
    \EIF[p(x|A=1)] & = \EIF\left[\frac{p(x, A=1)}{p(A=1)}\right]\\
    & = \EIF\left[\frac{\E[\mathbb{1}(X=x, A=1)]}{\E[A]}\right]\\
    & = \frac{1}{p(A=1)}\big[\mathbb{1}(X=x, A=1) - p(x, A=1)\\
    &\phantom{=} - p(x|A=1)[A - p(A=1)]\big]\\
    & = \frac{A}{p(A=1)}\big[\mathbb{1}(X=x) - p(x|A=1)\big].
\end{aligned}
\end{equation}
\textbf{Step 4: use Step 1 and Step 3 building blocks to compute $\EIF[\tau_1]$ and $\EIF[\tau_0]$}.
\begin{equation}
\begin{aligned}
    \EIF[\tau_1] & = \sum_x\Big[\frac{\mathbb{1}(X=x,A=1)}{p(x, 1)}\left[Y - \mu_1(x)\right]p(x|A=1)\\
    &\phantom{=}+\mu_1(x)\frac{A}{p(A=1)}\big[\mathbb{1}(X=x) - p(x|A=1)\big]\Big]\\
    &= \frac{A}{p(A=1)}(Y-\tau_1),\\
    \EIF[\tau_0] & = \sum_x\Big[\frac{\mathbb{1}(X=x,A=0)}{p(x, 0)}\left[Y - \mu_0(x)\right]p(x|A=1)\\
    &\phantom{=}+\mu_0(x)\frac{A}{p(A=1)}\big[\mathbb{1}(X=x) - p(x|A=1)\big]\Big]\\
    &= \frac{1-A}{p(A=1)}\frac{e(X)}{1-e(X)}(Y-\mu_0(X)) + \frac{A}{p(A=1)}(\mu_0(X)-\tau_0),
\end{aligned}
\end{equation}
\textbf{Step 5: final result}. Combining results from the previous step, we get:
\begin{equation}
\begin{aligned}
    \EIF[\tau_{ATT}] & = \frac{A}{p_1}[Y - \mu_0(X) - \tau_{ATT}] - \frac{1-A}{p_1}\frac{e(X)}{1-e(X)}[Y - \mu_0(X)],
\end{aligned}
\end{equation}
where $p_1=p(A=1)$ and $e(X)=p(A=1|X)$ is the propensity score. 
\section{Asymptotic Variance of the AIPW Estimator and Efficiency Bounds}
\label{app:variance}
In this section, we derive an expression for the variance of the ATT efficient influence function.
That corresponds to the semiparametric efficiency bound for ATT and it is attainable asymptotically by the AIPW estimator (and other doubly robust estimators) when both outcome and propensity models are well-specified and converge fast enough that the second-order error is negligible. We will use the efficient influence function derived in App.\,\ref{app:eif} and write it as $\phi_{ATT}(Z)$, where $Z=(A, X, Y)$ is the set of random variables of interest. We can split its variance into two terms:
\begin{equation}
    \V[\phi_{ATT}(Z)] = \V[\phi_1]+\V[\phi_0],
\end{equation}
where 
\begin{equation}
\begin{aligned}
    \phi_1 &= \frac{A}{p_1}[Y - \mu_0(X) - \tau_{ATT}],\\
    \phi_0 &= \frac{(1-A)}{p_1}\frac{e(X)}{1-e(X)}[Y - \mu_0(X)],
\end{aligned}    
\end{equation}
There is no covariance term, since $E[\phi_1\phi_0] = E[\phi_1] = E[\phi_0] = 0$. Let us start by analyzing the treated contribution:
%From the law of total variance and given that $E[Y - \mu_0(X) - \tau_{ATT} | A=1] = 0$, we get:
\begin{equation}
\begin{aligned}
\V[\phi_1] &= \frac{1}{p_1}\,\V[Y - \mu_0(X) | A=1]\\
&=\frac{1}{p_1}\Bigg[\E\Big[\V[Y - \mu_0(X) | X, A=1] \Big| A=1\Big]\\
&\phantom{=}+ \V\Big[\E[Y-\mu_0|X, A=1] \Big| A=1\Big] \Bigg]\\
& = \frac{1}{p_1}\Big[\E[\kappa^2_1(X) | A=1] + \V[\tau_1(X)|A=1]\Big],
\end{aligned}    
\end{equation}
where $\kappa^2_1(X)= \V[Y | X, A=1]$ is the outcome conditional variance on the treated and $\tau_1(X) = \mu_1(X) - \mu_0(X) | A=1$ is the conditional treatment effect on the treated. In the first equality, we have conditioned on $A=1$ and used the law of total variance. In the second equality, we have conditioned on $X$ and used again the law of total variance. Following similar steps for the historical control contribution, we get: 
\begin{equation}
\begin{aligned}
\V[\phi_0] &= \frac{1-p_1}{p_1^2}\,\V[w(X)(Y - \mu_0(X)) | A=0]\\
& =\frac{1-p_1}{p_1^2}\Bigg[\E\Big[w^2(X)\V[Y - \mu_0(X) | X, A=0] \Big| A=0\Big] \Bigg]\\
&=\frac{1-p_1}{p_1^2}\,\E\Big[w^2(X)\kappa^2_{0}(X)| A=0\Big],
\end{aligned}    
\end{equation}
where $w(X) = e(X) / (1 - e(X))$ are the propensity odds and $\kappa^2_{0}(X)=\V[Y | X, A=0]$ is the outcome conditional variance on the historical controls. In the first equality, we have conditioned on $A=0$ and used the law of total variance. In the second equality, we have conditioned on $X$ and used again the law of total variance. Note that, while both $Y$ and $w(X)$ are random variables, the propensity odds are random only through $X$, unlike $Y$ which has additional noise. Thus, in the second equality, propensity odds can be pulled out of the conditional variance as fixed numbers. Finally, combining both results we get:
\begin{equation}
% \begin{equation}
\label{eq:att}
\begin{aligned}
\V[\phi_{ATT}(Z)] &= \frac{1}{p_1}\,\E[\kappa^2_{1}(X) | A=1] +\frac{1-p_1}{p_1^2}\,\E[w^2(X)\kappa^2_{0}(X)| A=0] + \frac{1}{p_1}\,\V[\tau_1(X)|A=1].
\end{aligned}    
\end{equation}
\paragraph{Assuming Homogeneity and Homoskedasticity.}
We can simplify the expression for the asymptotic variance if we make additional assumptions. If we assume constant treatment effect, 
\begin{equation}
    \V[\tau_1(X)|A=1] = 0.
\end{equation}
Additionally, if we assume that residuals are homoskedastic the variance terms become independent of $X$ and we get:
\begin{equation}
\begin{aligned}
\V[\phi_{ATT}(Z)] &= \frac{\kappa^2_{1}}{p_1}+\frac{(1-p_1)\kappa^2_{0}}{p_1^2}\,\E[w^2(X)| A=0].\\
\end{aligned}    
\end{equation}
% \frac{(\E[w(X)\mid A=0])^2}{\E[w^2(X)\mid A=0]}
If we define 
\begin{equation}
    \gamma=\left(\frac{p_1}{1-p_1}\right)^2\frac{1}{\E[w^2(X)\mid A=0]},
\end{equation}
we can  rewrite the efficiency bound as:
\begin{equation}
\begin{aligned}
\V[\phi_{ATT}(Z)] &= \frac{\kappa^2_{1}}{p_1}+\frac{\kappa^2_{0}}{\gamma(1-p_1)}.
\end{aligned}    
\end{equation}
Finally, if we assume the same conditional variance in the treated and control populations $\kappa_1^2=\kappa_2^2=\kappa^2$, on a large sample of size $n$, the efficiency bound reduces to: 
\begin{equation}
\label{eq:att_hh}
\begin{aligned}
\V[\widehat{\tau}_{ATT}^{\mathrm{AIPW}}] = \frac{\V[\phi_{ATT}(Z)]}{n}&= \kappa^2\left(\frac{1}{n_1}+\frac{1}{n_0^{\mathrm{eff}}}\right).
\end{aligned}    
\end{equation}
In the last equality, we have defined the effective control sample size as $n_0^{\mathrm{eff}}=\gamma n_0$. As we already derived for the IPW estimator in \Eq{eq:gamma}, the efficiency factor $\gamma$ can be written as
\begin{equation}
\gamma \;=\; \left(\frac{p_1}{1-p_1}\right)^{2}\frac{1}{
\E\!\left[w^2(X) \mid A=0\right]}=\frac{1}{1+\chi^2\left(P_1(X)||P_0(X)\right)}
       \;\leq\; 1,
\end{equation}
where in the last equality we used the $\chi^2$ divergence between the baseline covariate distributions $P_{0,1}(X)=P(X|A=0,1)$. The conditional variance can also be rewritten, e.g., as $\kappa^2=\V[Y\mid X, A=1] = \V[Y\mid X, A=0] = \sigma_0^2(1-\rho_0^2)$ in terms of the marginal variance $\sigma_0^2$ and the correlation $\rho_0=\mathrm{Corr}[\mu_0(x), Y\mid A=0]$, assuming a constant treatment effect.

\section{Case Study Datasets}
\label{app:external-data}
In this appendix, we describe the datasets used for the case studies analysis of \Sec{sec:case-studies}. For each indication, we report the baseline characteristics of the trial analyzed as well as the baseline characteristics of the training and historical cohorts. The \emph{training} cohort is the full external pool used to train the outcome model; the \emph{historical} cohort is the subset that satisfies the trial eligibility criteria and is used in the propensity model and as the historical-control sample in the PSM, IPW, and AIPW estimators. We also list the per-endpoint standard deviations on the historical cohort that are used to standardize the bias reported in \Sec{sec:case-studies}.

\paragraph{Amyotrophic Lateral Sclerosis (Celebrex).}
The ALS analysis uses an aggregation of historical ALS clinical trials and observational studies to construct external controls for the Celebrex study. The training dataset corresponds to the full external pool. The historical dataset is the subset that satisfies the trial eligibility criteria, which we simplified to: age $\geq 18$ years, FVC $\geq 60\%$ predicted at baseline, and time since diagnosis within 5 years. The training cohort contains 13,496 subjects;
the historical (IE-eligible) cohort contains 4,643
subjects.

Tab.\,\ref{tab:als-trial-baseline} reports the baseline characteristics
of the trial cohort, split by randomized arm.
Tab.\,\ref{tab:als-baseline} summarizes the same characteristics for
the training and historical cohorts.
Tab.\,\ref{tab:als-sds} reports the
per-endpoint outcome SDs on the historical cohort used to compute the
absolute standardized bias.

\paragraph{Huntington's Disease (2CARE).}
The HD analysis uses an aggregation of historical HD natural-history and interventional studies to construct external controls for the 2CARE study. The training dataset corresponds to the full external pool. The historical dataset is the subset that satisfies the trial eligibility criteria, which we simplified to: age $\geq 16$ years, baseline UHDRS TFC $\geq 10$, and a manifest-HD or genetically confirmed HD diagnosis consistent with the 2CARE inclusion criteria (CAG repeats $\geq 36$ or manifest HD with a parent affected). The training cohort contains 19,446 subjects;
the historical (IE-eligible) cohort contains 12,775
subjects.

Tab.\,\ref{tab:hd-trial-baseline} reports the baseline characteristics
of the trial cohort, split by randomized arm.
Tab.\,\ref{tab:hd-baseline} summarizes the same characteristics for
the training and historical cohorts.
Tab.\,\ref{tab:hd-sds} reports the
per-endpoint outcome SDs on the historical cohort used to compute the
absolute standardized bias.

\begin{table}
    \centering
    \small
    \renewcommand{\arraystretch}{1.05}
    \caption{Baseline characteristics of the Celebrex study data. Continuous variables are reported as
    mean $\pm$ standard deviation; binary variables as count (percentage).
    Statistics are computed on non-missing values; the rightmost column
    reports the percentage of subjects with a missing baseline value in
    each arm. Diagnostic delay is defined as the absolute difference between symptom onset day and diagnosis day. ALSFRS-R pre-slope is defined as (48 - ALSFRS-R score at baseline) / (days since symptom onset). FVC \% predicted 
    is the measured FVC at baseline, expressed as a percentage of the value
    predicted for a healthy reference subject with the same age, sex, and height. FVC \% pre-slope is computed as (100 - FVC \% at baseline) / (days since symptom onset). Site of onset was not available.}
    \label{tab:als-trial-baseline}
    \begin{tabular}{p{6.2cm} c c c}
        \toprule
        & \textbf{Control} & \textbf{Treated} & \textbf{Missing} \\
        & ($N=99$) & ($N=201$) & $\%_{\mathrm{ctrl}}/\%_{\mathrm{trt}}$ \\
        \midrule
        Age (years) & 55.02 $\pm$ 12.40 & 54.54 $\pm$ 11.79 & 0.0\% / 0.0\% \\
        Sex Female & 33 (33.3\%) & 73 (36.3\%) & 0.0\% / 0.0\% \\
        Race Caucasian & 89 (89.9\%) & 178 (88.6\%) & 0.0\% / 0.0\% \\
        BMI (kg/m$^2$) & 26.83 $\pm$ 4.41 & 26.60 $\pm$ 4.39 & 1.0\% / 0.5\% \\
        Weight (kg) & 79.63 $\pm$ 14.73 & 78.56 $\pm$ 15.87 & 1.0\% / 0.0\% \\
        Height (cm) & 172.07 $\pm$ 9.50 & 171.30 $\pm$ 9.23 & 0.0\% / 0.5\% \\
        Days since diagnosis & -313.53 $\pm$ 327.36 & -294.40 $\pm$ 286.38 & 0.0\% / 0.0\% \\
        Days since symptom onset & -708.40 $\pm$ 448.12 & -689.24 $\pm$ 425.93 & 0.0\% / 0.0\% \\
        Diagnostic delay (days) & 394.88 $\pm$ 335.62 & 394.84 $\pm$ 318.08 & 0.0\% / 0.0\% \\
        ALSFRS-R total score & 39.42 $\pm$ 5.03 & 39.20 $\pm$ 5.24 & 0.0\% / 0.5\% \\
        ALSFRS-R pre-slope (units/day) & -0.02 $\pm$ 0.01 & -0.02 $\pm$ 0.02 & 0.0\% / 0.5\% \\
        FVC \% predicted & 79.10 $\pm$ 14.03 & 81.65 $\pm$ 15.83 & 0.0\% / 0.0\% \\
        FVC \% pre-slope (units/day) & -0.05 $\pm$ 0.03 & -0.05 $\pm$ 0.04 & 8.1\% / 14.4\% \\
        El Escorial Definite & 39 (39.4\%) & 75 (37.3\%) & 0.0\% / 0.0\% \\
        El Escorial Probable & 43 (43.4\%) & 84 (41.8\%) & 0.0\% / 0.0\% \\
        El Escorial Probable (lab-supported) & 15 (15.2\%) & 38 (18.9\%) & 0.0\% / 0.0\% \\
        El Escorial Possible & 0 (0.0\%) & 0 (0.0\%) & 0.0\% / 0.0\% \\
        El Escorial Suspected & 2 (2.0\%) & 4 (2.0\%) & 0.0\% / 0.0\% \\
        Family history of ALS & 10 (10.1\%) & 9 (4.5\%) & 0.0\% / 0.0\% \\
        Known genetic mutation & 0 (0.0\%) & 0 (0.0\%) & 0.0\% / 0.0\% \\
        Non-invasive ventilation at baseline & 99 (100.0\%) & 201 (100.0\%) & 0.0\% / 0.0\% \\
        Taking riluzole & 4 (4.0\%) & 6 (3.0\%) & 0.0\% / 0.0\% \\
        Taking edaravone & 0 (0.0\%) & 0 (0.0\%) & 0.0\% / 0.0\% \\
        \bottomrule
    \end{tabular}
\end{table}

\begin{table}
    \centering
    \small
    \renewcommand{\arraystretch}{1.05}
    \caption{Baseline characteristics of the training and historical cohorts
    for the Celebrex study analysis. 
    The training cohort is the full
    external pool used for outcome model training; the historical cohort is
    the subset of the training cohort that satisfies the trial eligibility
    criteria, used in the PSM, IPW, and AIPW estimators. Continuous variables
    are reported as mean $\pm$ standard deviation; binary variables as count
    (percentage). Statistics are computed on non-missing values; the
    rightmost column reports the percentage of subjects with a missing
    baseline value in the training and historical cohorts, respectively. See Tab.\,\ref{tab:als-trial-baseline} for a definition of pre-slopes and FVC \% predicted.}
    \label{tab:als-baseline}
    \begin{tabular}{p{6.2cm} c c c}
        \toprule
        & \textbf{Training} & \textbf{Historical} & \textbf{Missing} \\
        & ($N=13496$) & ($N=4643$) & $\%_{\mathrm{train}}/\%_{\mathrm{hist}}$ \\
        \midrule
        Age (years) & 58.34 $\pm$ 11.92 & 56.90 $\pm$ 11.57 & 2.4\% / 0.0\% \\
        Sex Female & 5270 (39.1\%) & 1662 (35.8\%) & 0.0\% / 0.0\% \\
        Race Caucasian & 8569 (95.3\%) & 4348 (95.2\%) & 33.4\% / 1.6\% \\
        BMI (kg/m$^2$) & 26.10 $\pm$ 4.77 & 26.88 $\pm$ 4.79 & 58.8\% / 55.8\% \\
        Weight (kg) & 76.25 $\pm$ 16.55 & 79.03 $\pm$ 16.54 & 46.8\% / 45.4\% \\
        Height (cm) & 171.15 $\pm$ 9.94 & 172.14 $\pm$ 10.05 & 50.8\% / 55.7\% \\
        Days since diagnosis & -334.17 $\pm$ 417.79 & -287.07 $\pm$ 294.75 & 53.8\% / 0.0\% \\
        Days since symptom onset & -719.06 $\pm$ 558.83 & -663.73 $\pm$ 447.41 & 18.3\% / 0.7\% \\
        Diagnostic delay (days) & 390.36 $\pm$ 347.96 & 379.60 $\pm$ 330.65 & 54.5\% / 0.9\% \\
        ALSFRS-R total score & 36.77 $\pm$ 7.12 & 37.62 $\pm$ 5.41 & 25.7\% / 14.7\% \\
        ALSFRS-R pre-slope (units/day) & -0.02 $\pm$ 0.11 & -0.02 $\pm$ 0.02 & 31.4\% / 15.4\% \\
        FVC \% predicted & 78.49 $\pm$ 23.32 & 86.95 $\pm$ 15.77 & 49.9\% / 35.2\% \\
        FVC \% pre-slope (units/day) & -0.05 $\pm$ 0.05 & -0.04 $\pm$ 0.04 & 62.5\% / 48.3\% \\
        El Escorial Definite & 1485 (21.1\%) & 1151 (34.7\%) & 47.9\% / 28.5\% \\
        El Escorial Probable & 1401 (19.9\%) & 1156 (34.8\%) & 47.9\% / 28.5\% \\
        El Escorial Probable (lab-supported) & 847 (12.1\%) & 657 (19.8\%) & 47.9\% / 28.5\% \\
        El Escorial Possible & 3137 (44.6\%) & 277 (8.3\%) & 47.9\% / 28.5\% \\
        El Escorial Suspected & 157 (2.2\%) & 77 (2.3\%) & 47.9\% / 28.5\% \\
        Site of onset Bulbar & 2471 (21.8\%) & 835 (18.1\%) & 16.0\% / 0.6\% \\
        Site of onset Limb & 7567 (66.5\%) & 3179 (68.5\%) & 15.7\% / 0.0\% \\
        Site of onset Other & 585 (5.4\%) & 357 (8.4\%) & 20.1\% / 8.0\% \\
        Family history of ALS & 439 (18.8\%) & 189 (14.0\%) & 82.7\% / 70.9\% \\
        Known genetic mutation & 264 (2.0\%) & 47 (1.0\%) & 0.0\% / 0.0\% \\
        Non-invasive ventilation at baseline & 3022 (22.4\%) & 203 (4.4\%) & 0.0\% / 0.0\% \\
        Taking riluzole & 4229 (39.6\%) & 2149 (46.3\%) & 21.0\% / 0.0\% \\
        Taking edaravone & 79 (0.7\%) & 60 (1.3\%) & 21.0\% / 0.0\% \\
        \bottomrule
    \end{tabular}
\end{table}

\begin{table}
    \centering
    \small
    \renewcommand{\arraystretch}{1.15}
    \caption{Outcome standard deviations used to compute the absolute
    standardized bias the Celebrex analysis. For each endpoint, the standard deviation
    is computed on the historical (IE-eligible) cohort using all subjects
    with a non-missing value at the analysis time.}
    \label{tab:als-sds}
    \begin{tabular}{p{8cm} c c}
        \toprule
        \textbf{Endpoint (change from baseline at 12 months)} & $n_{\mathrm{hist}}$ & \textbf{SD} \\
        \midrule
        ALSFRS-R total  & 1103 & 7.46 \\
        FVC \% predicted  & 437 & 21.03 \\
        \bottomrule
    \end{tabular}
\end{table}

\begin{table}
    \centering
    \small
    \renewcommand{\arraystretch}{1.05}
    \caption{Baseline characteristics of the 2CARE study data. Continuous variables are reported as
    mean $\pm$ standard deviation; binary variables as count (percentage).
    Statistics are computed on non-missing values; the rightmost column
    reports the percentage of subjects with a missing baseline value in
    each arm.}
    \label{tab:hd-trial-baseline}
    \begin{tabular}{p{6.2cm} c c c}
        \toprule
        & \textbf{Control} & \textbf{Treated} & \textbf{Missing} \\
        & ($N=303$) & ($N=297$) & $\%_{\mathrm{ctrl}}/\%_{\mathrm{trt}}$ \\
        \midrule
        Age (years) & 50.24 $\pm$ 11.53 & 50.21 $\pm$ 11.89 & 0.0\% / 0.3\% \\
        Sex Female & 161 (53.1\%) & 144 (48.5\%) & 0.0\% / 0.0\% \\
        Ethnicity Hispanic & 7 (2.3\%) & 5 (1.7\%) & 1.0\% / 0.0\% \\
        Education (years) & 14.28 $\pm$ 2.29 & 14.13 $\pm$ 2.30 & 0.0\% / 0.0\% \\
        CAG repeats & 43.91 $\pm$ 3.83 & 44.05 $\pm$ 4.16 & 0.0\% / 0.0\% \\
        CAP score & 102.55 $\pm$ 15.05 & 102.61 $\pm$ 14.94 & 0.0\% / 0.3\% \\
        PIN score & 2.63 $\pm$ 1.35 & 2.70 $\pm$ 1.30 & 2.6\% / 1.7\% \\
        cUHDRS & 10.58 $\pm$ 2.88 & 10.28 $\pm$ 2.86 & 2.6\% / 1.7\% \\
        Age at HD diagnosis & 47.35 $\pm$ 11.43 & 47.31 $\pm$ 12.08 & 3.0\% / 2.0\% \\
        Age at symptom onset & 43.31 $\pm$ 11.37 & 43.74 $\pm$ 11.78 & 3.3\% / 0.7\% \\
        Motor symptoms & 217 (74.1\%) & 236 (80.0\%) & 3.3\% / 0.7\% \\
        Cognitive symptoms & 62 (21.2\%) & 72 (24.4\%) & 3.3\% / 0.7\% \\
        Depressive symptoms & 169 (55.8\%) & 183 (61.6\%) & 0.0\% / 0.0\% \\
        Obsessive symptoms & 33 (10.9\%) & 23 (7.7\%) & 0.0\% / 0.0\% \\
        Psychotic symptoms & 10 (3.3\%) & 4 (1.3\%) & 0.0\% / 0.0\% \\
        Father with HD & 128 (45.7\%) & 122 (46.4\%) & 7.6\% / 11.4\% \\
        Mother with HD & 145 (51.2\%) & 133 (49.3\%) & 6.6\% / 9.1\% \\
        History of substance abuse & 40 (13.2\%) & 40 (13.5\%) & 0.0\% / 0.0\% \\
        UHDRS TFC total & 11.02 $\pm$ 1.46 & 10.78 $\pm$ 1.51 & 1.0\% / 0.7\% \\
        UHDRS Functional Assessment & 22.88 $\pm$ 2.21 & 22.67 $\pm$ 2.28 & 1.3\% / 1.0\% \\
        UHDRS Independence & 90.00 $\pm$ 8.81 & 89.07 $\pm$ 9.13 & 1.0\% / 0.7\% \\
        UHDRS Total Motor Score & 27.52 $\pm$ 13.90 & 28.08 $\pm$ 13.33 & 2.0\% / 0.7\% \\
        UHDRS Motor: chorea & 8.55 $\pm$ 4.79 & 8.69 $\pm$ 4.25 & 1.0\% / 0.7\% \\
        SDMT (correct) & 29.91 $\pm$ 12.11 & 29.03 $\pm$ 11.43 & 1.7\% / 1.3\% \\
        Letter fluency (correct) & 26.52 $\pm$ 12.05 & 25.55 $\pm$ 11.53 & 1.0\% / 1.0\% \\
        Stroop color (correct) & 51.44 $\pm$ 16.45 & 51.85 $\pm$ 17.36 & 1.0\% / 1.3\% \\
        Stroop word (correct) & 65.01 $\pm$ 18.55 & 63.88 $\pm$ 20.21 & 1.0\% / 1.3\% \\
        HD-ISS stage & 2.78 $\pm$ 0.45 & 2.81 $\pm$ 0.43 & 4.6\% / 2.0\% \\
        Stroop interference (correct) & 29.54 $\pm$ 11.22 & 30.41 $\pm$ 10.92 & 1.3\% / 1.3\% \\
        Taking VMAT2 inhibitor & 18 (5.9\%) & 15 (5.1\%) & 0.0\% / 0.0\% \\
        Taking 2nd-gen antipsychotic & 50 (16.5\%) & 53 (17.8\%) & 0.0\% / 0.0\% \\
        \bottomrule
    \end{tabular}
\end{table}

\begin{table}
    \centering
    \small
    \renewcommand{\arraystretch}{1.05}
    \caption{Baseline characteristics of the training and historical cohorts
    for the 2CARE study analysis. 
    The training cohort is the full
    external pool used for outcome model training; the historical cohort is
    the subset of the training cohort that satisfies the trial eligibility
    criteria, used in the PSM, IPW, and AIPW estimators. Continuous variables
    are reported as mean $\pm$ standard deviation; binary variables as count
    (percentage). Statistics are computed on non-missing values; the
    rightmost column reports the percentage of subjects with a missing
    baseline value in the training and historical cohorts, respectively.}
    \label{tab:hd-baseline}
    \begin{tabular}{p{6.2cm} c c c}
        \toprule
        & \textbf{Training} & \textbf{Historical} & \textbf{Missing} \\
        & ($N=19446$) & ($N=12775$) & $\%_{\mathrm{train}}/\%_{\mathrm{hist}}$ \\
        \midrule
        Age (years) & 48.06 $\pm$ 13.47 & 44.87 $\pm$ 13.06 & 0.0\% / 0.0\% \\
        Sex Female & 10661 (54.8\%) & 7000 (54.8\%) & 0.0\% / 0.0\% \\
        Ethnicity Hispanic & 368 (1.9\%) & 243 (1.9\%) & 1.4\% / 1.3\% \\
        Education (years) & 12.28 $\pm$ 2.67 & 12.70 $\pm$ 2.41 & 6.6\% / 6.4\% \\
        CAG repeats & 43.17 $\pm$ 3.17 & 42.82 $\pm$ 2.97 & 1.7\% / 1.5\% \\
        CAP score & 94.34 $\pm$ 23.49 & 85.69 $\pm$ 21.78 & 1.7\% / 1.5\% \\
        PIN score & 1.74 $\pm$ 2.07 & 0.91 $\pm$ 1.67 & 15.6\% / 9.5\% \\
        cUHDRS & 11.72 $\pm$ 5.00 & 14.14 $\pm$ 3.35 & 15.2\% / 8.6\% \\
        Age at HD diagnosis & 49.05 $\pm$ 11.91 & 48.52 $\pm$ 11.83 & 29.4\% / 42.9\% \\
        Age at symptom onset & 45.81 $\pm$ 11.64 & 45.95 $\pm$ 11.62 & 31.7\% / 43.9\% \\
        Motor symptoms & 13889 (76.8\%) & 7618 (65.2\%) & 7.0\% / 8.6\% \\
        Cognitive symptoms & 8486 (46.9\%) & 3473 (29.7\%) & 7.0\% / 8.6\% \\
        Depressive symptoms & 13334 (72.2\%) & 8170 (68.6\%) & 5.1\% / 6.8\% \\
        Obsessive symptoms & 10052 (54.5\%) & 5411 (45.4\%) & 5.1\% / 6.8\% \\
        Psychotic symptoms & 2270 (12.3\%) & 879 (7.4\%) & 5.1\% / 6.8\% \\
        Father with HD & 7935 (45.7\%) & 5117 (45.0\%) & 10.7\% / 10.9\% \\
        Mother with HD & 8896 (50.6\%) & 6051 (52.5\%) & 9.5\% / 9.8\% \\
        History of substance abuse & 2923 (19.1\%) & 1943 (19.6\%) & 21.4\% / 22.2\% \\
        UHDRS TFC total & 10.21 $\pm$ 3.41 & 12.32 $\pm$ 1.03 & 0.9\% / 0.0\% \\
        UHDRS Functional Assessment & 21.13 $\pm$ 5.65 & 24.20 $\pm$ 1.46 & 4.1\% / 3.5\% \\
        UHDRS Independence & 86.89 $\pm$ 16.45 & 95.88 $\pm$ 7.01 & 2.0\% / 1.6\% \\
        UHDRS Total Motor Score & 22.79 $\pm$ 22.02 & 12.25 $\pm$ 13.69 & 1.0\% / 0.5\% \\
        UHDRS Motor: chorea & 5.58 $\pm$ 5.64 & 3.56 $\pm$ 4.51 & 0.6\% / 0.2\% \\
        SDMT (correct) & 34.69 $\pm$ 16.96 & 40.95 $\pm$ 15.23 & 13.4\% / 7.7\% \\
        Letter fluency (correct) & 27.95 $\pm$ 15.07 & 32.89 $\pm$ 14.19 & 28.1\% / 25.9\% \\
        Stroop color (correct) & 54.74 $\pm$ 21.19 & 63.23 $\pm$ 18.15 & 8.5\% / 5.7\% \\
        Stroop word (correct) & 72.00 $\pm$ 26.42 & 82.52 $\pm$ 22.09 & 7.5\% / 4.3\% \\
        Stroop interference (correct) & 31.83 $\pm$ 14.47 & 36.81 $\pm$ 13.09 & 16.1\% / 10.6\% \\
        HD-ISS stage & 2.27 $\pm$ 0.91 & 1.90 $\pm$ 0.91 & 8.2\% / 9.9\% \\
        Taking VMAT2 inhibitor & 1445 (9.4\%) & 365 (3.6\%) & 20.7\% / 21.5\% \\
        Taking 2nd-gen antipsychotic & 2563 (16.6\%) & 695 (6.9\%) & 20.7\% / 21.5\% \\
        \bottomrule
    \end{tabular}
\end{table}

\begin{table}
    \centering
    \small
    \renewcommand{\arraystretch}{1.15}
    \caption{Outcome standard deviations used to compute the absolute
    standardized bias for the 2CARE analysis. For each endpoint, the SD
    is computed on the historical (IE-eligible) cohort using all subjects
    with a non-missing value at the analysis time.}
    \label{tab:hd-sds}
    \begin{tabular}{p{8cm} c c}
        \toprule
        \textbf{Endpoint (change from baseline at 4 years)} & $n_{\mathrm{hist}}$ & \textbf{SD} \\
        \midrule
        UHDRS TFC  & 2396 & 2.35 \\
        UHDRS FA  & 2289 & 3.89 \\
        UHDRS Independence  & 2339 & 11.02 \\
        UHDRS TMS  & 2381 & 10.46 \\
        \bottomrule
    \end{tabular}
\end{table}

\section{Propensity Score and Digital Twin Models}
\label{app:digital-twin-models}
\subsection{Propensity Score Model}
\paragraph{Architecture.}
We estimate the propensity score $e(X) = P(A=1\mid X)$, where $A=1$
indicates trial membership and $A=0$ indicates historical control, with
an $\ell_2$-regularized logistic regression. The model is fit on the
pooled set of trial and historical-control subjects. The estimator is implemented as a pipeline consisting of (i) iterative imputation of missing
baseline features, run for five rounds with a Bayesian-ridge
estimator on the remaining features; (ii) standardization of each
feature; and (iii) logistic regression with inverse regularization strength $C$
selected from a logarithmic grid. The full pipeline, including the
imputer and scaler, is refit on each training fold so no held-out
information leaks through the preprocessing stages.

\paragraph{Training and hyperparameter selection.} We use nested stratified cross-validation: 
\begin{itemize}
    \item An outer 5-fold split produces out-of-fold predicted probabilities used in all relevant estimators. Folds are stratified on the trial-membership label.
    \item Within each outer training set, an inner 3-fold split is
    used to select $C$ by maximizing held-out log-likelihood. Folds are stratified on the trial-membership label.
\end{itemize}
As discussed in \Sec{sec:case-studies}, we restrict the analysis to the
common-support region of the estimated propensity distribution, defined
as the overlap of the trial and historical-control propensity ranges;
this stabilizes the propensity odds and ensures that the estimators are
evaluated where positivity is well-supported. The baseline features
used for the propensity models in each case study are reported in
Tabs.~\ref{tab:als-features} and \ref{tab:hd-features}.

\subsection{Digital Twin Model}
We trained independent models for each continuous outcome of interest.
Outcome models predict a participant's outcome trajectory from baseline
covariates and are parameterized through a per-participant
\emph{velocity} of change rather than predicting absolute outcome values
directly. Concretely, given baseline features $X_i$, baseline outcome
$y_i(t_0)$, and prediction time $t > t_0$, the predicted outcome is given by
\begin{equation}
\hat{y}_i(t) \;=\; y_i(t_0) \;+\; \hat{v}(X_i)\,(t - t_0),
\end{equation}
where $\hat{v}(\cdot)$ is a learned regressor of the empirical velocity
$v_i(t) = (y_i(t) - y_i(t_0)) / (t - t_0)$ on the baseline covariates
$X_i$. The velocity formulation approximates the progression of the endpoints of interest as linear in time and enforces that predictions collapse to
the observed baseline at $t = t_0$.

\paragraph{Architecture.} The velocity regressor is implemented as a
pipeline consisting of (i) iterative imputation of missing baseline
features and (ii) a random forest regressor. The iterative imputer fills each feature with missing values
in turn by regressing it on the remaining features and is run for a
fixed number of rounds; we use a Bayesian-ridge estimator
and five rounds. The imputer sits inside a cross-validation loop, so
it is refit on each training fold (and on each bootstrap resample).
The architecture is intentionally kept simple: because our variance
estimation procedure for some of the estimators requires retraining the
digital twin model on every bootstrap resample, a lightweight, fast-to-fit
model is preferable to a more computationally expensive alternative.
Random forests provide a good practical trade-off between predictive
accuracy and the runtime budget needed for repeated refitting across
bootstrap iterations. Participant trajectories provide multiple training
rows per participant (one per post-baseline visit); all rows from the
same participant are constrained to the same cross-validation fold to
prevent leakage. The goal of the case study analyses is not necessarily to use optimal digital twin models, but rather to explore the feasibility and potential advantages of digital-twin-based approaches compared with other methods. We expect that more expressive architectures, such as those described in \cite{alam2024digital}, could further improve results when properly optimized.

\paragraph{Training and hyperparameter selection.}
The model is fit using nested grouped cross-validation, with
groups defined by participant identifier:
\begin{itemize}
    \item An outer 5-fold split produces out-of-fold predictions used
    downstream as digital twins.
    \item Within each outer training set, an inner 3-fold split is
    used to select the random forest hyperparameters (number of
    estimators, maximum depth, and minimum samples per leaf),
    minimizing mean squared error on velocities.
\end{itemize}
For inference on participants that do not appear in the training set,
predictions are computed by averaging velocity predictions across the
five outer-fold models. The baseline features
used for the outcome models in each case study are reported in
Tabs.~\ref{tab:als-features} and \ref{tab:hd-features}.

\begin{table}
    \centering
    \small
    \renewcommand{\arraystretch}{1.15}
    \caption{Baseline features used in the propensity and outcome models
    for the Celebrex analysis. The propensity model is fit on the union of
    trial and historical-control subjects; the outcome model
    is fit on the training cohort. See Tabs.~\ref{tab:als-trial-baseline} and \ref{tab:als-baseline} for details about the datasets.}
    \label{tab:als-features}
    \begin{tabular}{p{2.3cm} p{12cm}}
        \toprule
        \textbf{Model} & \textbf{Features} \\
        \midrule
        Propensity & Age; Sex Female; El Escorial Definite; El Escorial Possible; Days since symptom onset; ALSFRS-R pre-slope; FVC \% pre-slope; ALSFRS-R total score; FVC \% predicted. \\
        \addlinespace
        Outcome & Age; Family history of ALS; ALSFRS-R bulbar subscore; ALSFRS-R fine-motor subscore; ALSFRS-R gross-motor subscore; ALSFRS-R respiratory subscore; ALSFRS-R total score; ALSFRS-R pre-slope; BMI; El Escorial Definite; El Escorial Possible; El Escorial Probable; El Escorial Probable lab-supported; El Escorial Suspected; Days since diagnosis; Diagnostic delay; FVC volume; FVC volume pre-slope; FVC \%; FVC \% pre-slope; Known genetic mutation; Height; Non-invasive ventilation at baseline; Non-ALS concomitant medications; Race Caucasian; Sex Female; Site of onset Bulbar; Site of onset Limb; Days since symptom onset; Taking amitriptyline; Taking baclofen; Taking edaravone; Taking riluzole; Taking sodium phenylbutyrate; Weight. \\
        \bottomrule
    \end{tabular}
\end{table}

\begin{table}
    \centering
    \small
    \renewcommand{\arraystretch}{1.15}
    \caption{Baseline features used in the propensity and outcome models
    for the 2CARE analysis. The propensity model is fit on the union of
    trial and historical-control subjects; the outcome model
    is fit on the training cohort. See Tabs.~\ref{tab:hd-trial-baseline} and \ref{tab:hd-baseline} for details about the datasets.}
    \label{tab:hd-features}
    \begin{tabular}{p{2.3cm} p{12cm}}
        \toprule
        \textbf{Model} & \textbf{Features} \\
        \midrule
        Propensity & Age; CAG repeats; CAP score; Sex Female; Cognitive symptoms; Motor symptoms; UHDRS Functional Assessment; UHDRS Independence; UHDRS TFC total; UHDRS Total Motor Score; UHDRS Motor chorea; PIN score; HD-ISS stage. \\
        \addlinespace
        Outcome & Age; CAG repeats; CAP score; cUHDRS; Age at HD diagnosis; Education; Ethnicity Hispanic; Family history age at HD onset; Father with HD; Mother with HD; History of suicidal ideation; History of substance abuse; Letter fluency; PIN score; SDMT; Sex Female; Stroop color; Stroop interference; Stroop word; Age at symptom onset; Symptom onset Cognitive; Symptom onset Motor; Symptom onset Other; Symptom onset Psychiatric; Cognitive symptoms; Depressive symptoms; Motor symptoms; Obsessive symptoms; Psychotic symptoms; Taking 1st-gen antipsychotic; Taking 1st-gen antipsychotic injectable; Taking 2nd-gen antipsychotic; Taking 2nd-gen antipsychotic injectable; Taking VMAT2 inhibitor; UHDRS Behavioral Assessment; UHDRS Functional Assessment; UHDRS Independence; UHDRS TFC total; UHDRS Motor chorea; UHDRS Total Motor Score. \\
        \bottomrule
    \end{tabular}
\end{table}

\end{document}